\documentclass{article}

\usepackage[final]{neurips_2025}

\usepackage[utf8]{inputenc} 
\usepackage[T1]{fontenc}    
\usepackage{hyperref}       
\usepackage{url}            
\usepackage{booktabs}       
\usepackage{amsfonts}       
\usepackage{nicefrac}       
\usepackage[table,dvipsnames]{xcolor}

\usepackage{microtype}
\usepackage{inconsolata}
\usepackage{float}
\usepackage{graphicx}
\usepackage{multirow}
\usepackage{array}
\usepackage{colortbl}
\usepackage{makecell}
\usepackage{amsmath}
\usepackage{amsfonts}
\usepackage{tikz}
\usepackage{textgreek}
\usepackage[dvipsnames]{xcolor}
\usepackage{pifont} 
\newcommand{\mscellf}{\rowcolor{Cerulean!12}}
\newcommand{\mscells}{\rowcolor{Cerulean!18}}
\newcommand{\mscellt}{\rowcolor{Cerulean!25}}
\newcommand{\mscellfo}{\rowcolor{Cerulean!30}}

\usepackage{algorithm}
\usepackage{algorithmicx}
\usepackage{algpseudocode}
\usepackage{wrapfig} 
\usepackage{caption}
\usepackage{subcaption}
\usepackage{tcolorbox}
\tcbuselibrary{listings, breakable}

\title{InfiniPot-V: Memory-Constrained KV Cache Compression for Streaming Video Understanding }

\author{%
  \setcounter{footnote}{1}
  Minsoo Kim$^{1}$\thanks{Work done during an internship at Qualcomm Technologies, Inc.}\quad
  Kyuhong Shim$^{2}$\quad
  Jungwook Choi$^{1}$\thanks{Corresponding authors.}\quad
  Simyung Chang$^{3}$\footnotemark[3] \\
  $^{1}$Hanyang University \: $^{2}$Sungkyunkwan University \\
  $^{3}$Qualcomm AI Research, Qualcomm Korea YH\thanks{Qualcomm AI Research, an initiative of Qualcomm Technologies, Inc.} \\
  \texttt{\small\{minsoo2333, choij\}@hanyang.ac.kr} \,
  \texttt{\small khshim@skku.edu} \\
  \texttt{\small simychan@qti.qualcomm.com}
}

\begin{document}

\maketitle

\begin{abstract}
Modern multimodal large language models (MLLMs) can reason over hour-long video, yet their key–value (KV) cache grows linearly with time—quickly exceeding the fixed memory of phones, AR glasses, and edge robots. Prior compression schemes either assume the whole video and user query are available offline or must first build the full cache, so memory still scales with stream length. InfiniPot-V is the first training-free, query-agnostic framework that enforces a hard, length-independent memory cap for \textit{streaming} video understanding. During video encoding it monitors the cache and, once a user-set threshold is reached, runs a lightweight compression pass that (i) removes temporally redundant tokens via Temporal-axis Redundancy (TaR) metric and (ii) keeps semantically significant tokens via Value-Norm (VaN) ranking. Across four open-source MLLMs and four long-video and streaming-video benchmarks, InfiniPot-V cuts peak GPU memory by up to 94\%, sustains real-time generation, and matches or surpasses full-cache accuracy—even in multi-turn dialogues. By dissolving the KV cache bottleneck without retraining or query knowledge, InfiniPot-V closes the gap for on-device streaming video assistants.
\end{abstract}

\section{Introduction}
\label{sec:intro}

Recent advances in multimodal large language models (MLLMs) have dramatically expanded the scope of visual reasoning.  Vision–language instruction tuning now allows a single backbone to answer open-ended questions over long video sequences \cite{llava,openai2023gpt4v,qwen2.5}, while context-extension techniques such as FlashAttention-2 and RingAttention push the effective window into the million-token regime \cite{dao2023flashattention2,ringattention,peng2024yarn}.  These breakthroughs underpin a new generation of {\em streaming video assistants} and \emph{humanoid robots} that promise continuous, real-time scene understanding on mobile phones, AR glasses and edge robots~\cite{deepmind2023astra,openai2023gpt4o,waisberg2024meta,MorganStanley2024FigureAI}.

Streaming video understanding (SVU) diverges from conventional offline video understanding (OVU). Offline models see the entire clip and user query before inference, so they can tailor every compression or retrieval step. In streaming, frames arrive incrementally and future queries are unknown, forcing all pre-query processing to be {\em query-agnostic}. In addition, device memory is fixed, yet the transformer emits hundreds of tokens per frame, so the key–value (KV) cache grows linearly. For example, a 15-min, 10 fps clip processed by LLaVA-Next-Video-7B already needs demands \(\sim\!100\)\,GB of KV storage, far beyond the tens of gigabytes available on mobile or robotic platforms \cite{zhang2024llavanext-video,Karumbunathan2022JetsonAGXOrin}.  

Prior work tackles long-video memory constraints at three stages (Fig.\ref{fig:background}). {\em Frame Sampling} ~\cite{feichtenhofer2019slowfastnetworksvideorecognition} drops frames before encoding, reducing memory but severely degrading temporal coverage and accuracy. {\em Input-Vision Compression (IVC)}~\cite{shen2024longvu,tao2024dycokedynamiccompressiontokens} prunes redundant vision tokens after encoding, lowering Prefill load but still requiring the full vision token set to be stored in memory. {\em KV cache Compression (KVC)}~\cite{li2024snapkv,fu2024headkv} selects query-relevant tokens after the Prefill step, offering the highest accuracy but only after materializing the full KV cache. The challenge intensifies in streaming scenarios: memory usage for Frame Sampling, IVC, and KVC grows almost linearly with video length, eventually exceeding device limits. KV cache offloading (e.g., ReKV~\cite{Shangzhe2025streaming}) expands memory space yet incurs costly data transfer, repeated for each query. Thus, no existing approach delivers the key property SVU needs: a \textit{length-independent} and \textit{query-agnostic} streaming video compression.

A natural approach to address memory constraints in streaming video is to exploit the strong \textit{spatiotemporal redundancy} of video streams. We introduce \textbf{InfiniPot-V}, the first framework specifically designed for memory-constrained SVU. InfiniPot-V is \textit{training-free}, \textit{query-agnostic}, and operates \textit{continuously} during inference. When the KV cache reaches a user-defined memory threshold \(M\), it performs an in-place compression that frees space for new frames while preserving the semantic essence of prior context. This compression is guided by two lightweight and complementary metrics. \textit{Temporal-axis Redundancy (TaR)} models Key embeddings as a 3D tensor over time and removes tokens with high cosine similarity to recent frames, effectively pruning static or repetitive content. \textit{Value-Norm Importance (VaN)} ranks the remaining tokens by the \(\ell_2\) norm of their Value vectors—a strong, model-agnostic proxy for semantic salience—and applies a layer-adaptive pooling strategy. This compression is highly efficient, adding negligible latency while strictly enforcing memory limits.

Extensive evaluation confirms the effectiveness of this design. Across four open-source MLLMs (3B and 7B) and six long-video benchmarks—covering both offline (VideoMME, EgoSchema, MLVU, LongVideoBench) and streaming (RVS-Ego/Movie, OVO-Bench, StreamingBench) tasks—InfiniPot-V reduces input context length usage to as low as 6K for 50K-token contexts, with accuracy matching or exceeding full-cache baselines. It maintains real-time performance at 14 frames per second with only 0.5\% compression overhead. Additionally, its query-agnostic nature offers clear benefits in multi-turn dialogue settings (Appendix.~\ref{appn:multi_turn}). By eliminating the KV cache bottleneck without retraining or query dependency, InfiniPot-V paves the way for practical, on-device multimodal assistants.
\section{Background}\label{sec:background}

We aim to deploy streaming video understanding (SVU) applications~\cite{zhang2024flashvstream,Shangzhe2025streaming} in memory-constrained environments. Unlike offline video understanding (OVU)~\cite{MLVU,fu2024videomme,wu2024longvideobench}, which assumes access to the entire video, SVU must process arbitrarily long video streams and answer questions at any time step using only the frames observed up to that point. Given a video stream $V_T := [v_1, v_2, \dots, v_T]$ with $T$ frames and a set of questions $Q = {q_1, q_2, \dots, q_N}$, SVU answers each question $q_i$ at time $t$ ($1 \le t \le T$) using only the observed frames $V_t := [v_1, v_2, \dots, v_t]$.

As SVU deals with unbounded video streams, memory-efficient processing is essential. In this section, we describe how multimodal large language models (MLLMs) handle long videos, review prior approaches to memory reduction in OVU, and analyze their limitations when applied to SVU. (See Appendix.~\ref{sec:related} for a detailed discussion of related work.)

\subsection{Preliminary: Offline Long Video Understanding}

\begin{figure*}[t!]
    \centerline{\includegraphics[width=0.95\textwidth]{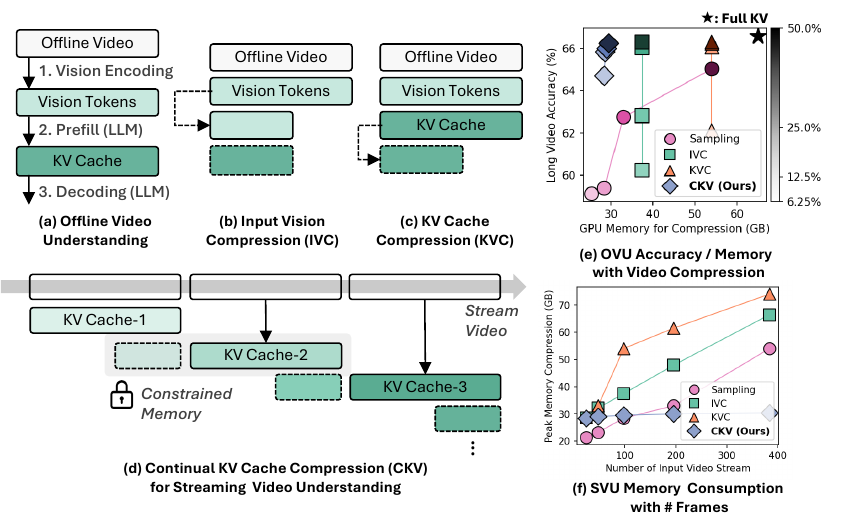}}\caption{\textbf{MLLMs Video Understanding and Compression.} (a) OVU pipeline; (b) IVC: compresses vision tokens after encoding; (c) KVC: compresses KV cache after prefill; (d) CKV: iteratively processes and compresses KV caches to constrain memory usage; (e) Accuracy vs. GPU memory consumption for compression across four token reduction ratios (50\%, 25\%, 12.5\%, 6.25\%) on MLVU using \texttt{Qwen-2-VL-7B}. LongVU\cite{shen2024longvu} is used for IVC, SnapKV\cite{li2024snapkv} for KVC; (f) GPU memory usage as input video stream length increases. IVC/KVC/CKV target a 6K cache; Sampling uses 1/4 of input frames. Measured on A100 80GB single GPU.}
    \vspace{-2mm}
\label{fig:background}

\end{figure*}

\paragraph{Video Processing in MLLMs.} Multimodal Large Language Models (MLLMs)~\cite{zhang2024llavanext-video,qwen2.5,Qwen2VL} process offline videos through a structured pipeline (Fig.~\ref{fig:background}(a)). Given a video $V_T := [v_1, v_2, \dots, v_T]$ of $T$ uniformly sampled frames, a vision encoder $f_{\text{ViT}}$ transforms each frame into visual tokens:
\begin{equation}
X = f_{\text{ViT}}(V_T) = [x_1, x_2, \dots, x_N] \in \mathbb{R}^{N \times D},
\label{eq:vision_encoding}
\end{equation}
where $N = P \times T$ denotes the total number of sampled tokens, where $P$ is the number of tokens per frame (determined by input resolution and ViT patch size), and $D$ is the token embedding dimension.

The token sequence $X$ is then passed to the LLM in two phases: Prefill and Decoding. During the Prefill phase (Fig.~\ref{fig:background}(a), step 2), all tokens are processed at once to construct the initial KV cache. The attention operation computes:
\begin{equation}
Q = XW_q, \quad K = XW_k, \quad V = XW_v, \quad O^{\text{attn}} = \text{Softmax}\left(\frac{QK^\top}{\sqrt{D}} + M\right)V,
\label{eq:attention_combined}
\end{equation}
where $W_q, W_k, W_v \in \mathbb{R}^{D \times D}$ are projection matrices and $M$ is a causal mask enforcing autoregressive decoding.

In the Decoding phase (Fig.~\ref{fig:background}(a), step 3), the model generates tokens one at a time using cached keys and values from the prefill phase. To avoid redundant computation, the KV cache $\mathcal{C} = (K, V)$ is updated incrementally:
\begin{equation}
\mathcal{C}_{t+1} = \left(\{K,k_{t+1}\}, \{V,v_{t+1}\}\right),
\label{eq:kv_cache}
\end{equation}
where $k_{t+1}, v_{t+1}$ correspond to the KV embeddings of the newly processed token.

\subsection{Offline Long-Video Compression Strategies}
\label{sec:motivation:offline}

Long videos produce extremely long token sequences $X$, leading to prohibitive GPU memory and latency during decoding.  Prior works tackle this bottleneck in the offline setting through three classes of methods (Fig.~\ref{fig:background}a–c):

\textbf{(1) Frame Sampling~\cite{feichtenhofer2019slowfastnetworksvideorecognition}.}  
Uniformly sampling a shorter clip $V'_{T'} \!\subseteq\! V_T$ reduces the input length and, hence, memory usage is also reduced proportional to compression rate. 

\textbf{(2) Input-Vision Compression (IVC)~\cite{shen2024longvu,tao2024dycokedynamiccompressiontokens}.}  
After vision encoding, IVC aggressively prunes redundant vision tokens, keeping only a salient subset $X' \!\subseteq\! X$ (Fig.~\ref{fig:background}b) to shrink the context fed into the language decoder for memory-compressed Prefill.

\textbf{(3) KV cache Compression (KVC)~\cite{li2024snapkv,fu2024headkv,chen2024image}.}  
Conduct compression after prefill: KVC computes importance scores  
$u_t\!=\!\sum_{i=N-w}^{N}\!\text{Attn}(x_i\!\rightarrow\!x_t)$  
over the last $w$ tokens and retain top-$M$ entries for the memory budget $M$ by applying eviction policy $\pi$, yielding a compressed cache  
$\mathcal{C}'=\pi(\mathcal{C})$ for memory-compressed Decoding.
(Fig.~\ref{fig:background}c). Note that the $\pi$ eviction policy is highly dependent on the content of the last $w$ tokens, reflecting the user query, and is thus referred to as query-dependent cache compression method (see Appendix.~\ref{appn:query_agnostic} for further analysis).

These techniques are effective when the entire video is available upfront, but they implicitly assume (i) unconstrained memory for compression and (ii) a known or easily approximated query.

\subsection{Challenges in Streaming Video Understanding}
\label{sec:motivation:streaming}

Fig.~\ref{fig:background}(e) compares three \textit{offline} compression methods on a fixed 50\,K-token video at four compression ratios (darker shades indicating higher ratios: 50\%, 25\%, 12.5\%, 6.25\%), revealing a fundamental trade-off between memory usage and accuracy. \textit{Frame sampling} skips frames to save memory, but severely degrades recognition accuracy. Increasing the sample ratio improves accuracy but quickly inflates memory usage. \textit{IVC} starts with a large memory footprint for all vision tokens before selecting which to discard. \textit{KVC}, which operates on more expressive key–value features, achieves the highest accuracy but requires the largest Prefill cache. Notably, even under a favorable offline setting—with full video access and an offline query—none of the methods achieve both high accuracy and low memory usage.

This trade-off becomes more severe in the streaming video understanding (SVU) setting. As shown in Fig.\ref{fig:background}(f), peak GPU memory usage increases with stream length. KVC exhibits near-linear memory growth, as it must materialize all vision tokens and build the full KV cache before compression. Furthermore, due to its query-dependent nature, KVC must \textit{re-execute} the memory-intensive prefill stage whenever the user query changes. Frame sampling and IVC also grow linearly, albeit more slowly, eventually exceeding the memory capacity of practical edge devices (e.g., 32GB\cite{Karumbunathan2022JetsonAGXOrin}) as the stream continues. ReKV~\cite{Shangzhe2025streaming}, a recent KVC method, addresses this by offloading the KV cache to CPU memory, but this introduces substantial offloading overhead and compression latency. 

These findings highlight two core requirements for SVU: (1) a fixed memory budget that does not grow with stream length, and (2) query-agnostic token retention strategies. Existing methods fail to meet at least one of these, limiting their suitability for SVU. 
To overcome this, we propose Continual KV cache compression (CKV), illustrated in Fig.~\ref{fig:background}(d). CKV processes frames in small blocks and compresses the cache whenever the fixed memory limit is reached, ensuring constant memory usage throughout streaming. Additionally, for query-agnostic token retention, our approach employs lightweight spatiotemporal metrics to identify and preserve semantically significant tokens without relying on future queries. As a result, despite operating under strict memory constraints, CKV achieves accuracy on par with or better than KVC (Fig.\ref{fig:background}(e)), while consuming far less memory than IVC or frame sampling (Fig.~\ref{fig:background}(f)). The algorithmic details are described in Sec.\ref{sec:method}.
\begin{algorithm}[h]
\caption{\small Continual KV cache Compression (CKV) with InfiniPot-V}
\label{alg:infinipot_v_main}
\footnotesize
\begin{algorithmic}[0]
\Require Memory budget $|M|$, target cache size $|C|$, TaR ratio $\alpha$
\State Initialize $K, V \gets \emptyset$ \Comment{Empty KV cache}
\While{video stream continues}
    \State (1) Process: $K_{\text{new}}, V_{\text{new}} \gets \text{Process new frame}$; $K \gets [K; K_{\text{new}}]$, $V \gets [V; V_{\text{new}}]$ \Comment{Append new tokens}
    \If{$\text{len}(K) \geq |M|$} \Comment{Memory budget exceeded}
        \State (2) Extract: $K_{\text{recent}}, V_{\text{recent}} \gets \text{recent }r\text{ frames from }K, V$ \Comment{$\text{len}(K) = \text{len}(V) = |M|$}
        \State (3) TaR: $s^{\text{TaR}} \gets \text{ComputeTaRScores}(K)$; $\mathcal{I}_{\text{TaR}} \gets \text{TopK}(s^{\text{TaR}}, \alpha|C| - \text{len}(K_{\text{recent}}))$ \Comment{Sec.~\ref{subsec:tar}}
        \State (4) VaN: $s^{\text{VaN}} \gets \text{ComputeAdaptiveVaNScores}(V)$; $\mathcal{I}_{\text{VaN}} \gets \text{TopK}(s^{\text{VaN}}, (1-\alpha)|C|)$ \Comment{Sec.~\ref{subsec:van}}
        \State (5) Combine: $\mathcal{I} \gets \mathcal{I}_{\text{TaR}} \cup \mathcal{I}_{\text{VaN}} \cup \text{Indices}(K_{\text{recent}})$; $K \gets K[\mathcal{I}]$, $V \gets V[\mathcal{I}]$ \Comment{Compress to $|C|$ size}
    \EndIf
    \If{user query arrives} \State Generate response using current $K, V$ \EndIf
\EndWhile
\end{algorithmic}
\end{algorithm}

\section{InfiniPot-V: Memory-Constrained Streaming Video Understanding}
\label{sec:method}

\begin{figure*}[t!]
\centerline{\includegraphics[width=0.9\textwidth]{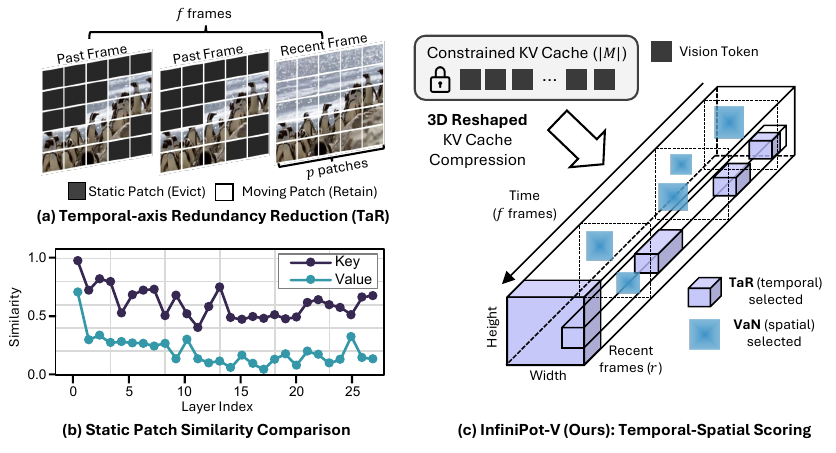}}
\caption{\textbf{Spatio-Temporal KV cache Compression (TaR and VaN).} (a) Temporal redundancy across adjacent frames, showing static patches that can be evicted from past frames; (b) Layer-wise cosine similarity of Key/Value embeddings for static patches between consecutive frames in \texttt{LLaVA-Next-Video-7B}; (c) InfiniPot-V performs query-agnostic spatiotemporal compression, reducing temporal redundancy with TaR and selecting tokens via VaN spatial scoring.}
\vspace{-0.1in}
\label{fig:cache_compression}
\end{figure*}

We present \textbf{InfiniPot-V}, a CKV framework designed for memory-constrained SVU. As shown in Fig.~\ref{fig:background}(d) and Algorithm~\ref{alg:infinipot_v_main}, InfiniPot-V processes video streams by applying continual KV cache compression within a fixed memory budget. 
In this framework, KV embeddings from incoming frames are stored until the memory limit $|M|$ is reached. At that point, compression reduces the cache to a smaller target size $|C|$ ($|M| \gg |C|$), retaining only the most essential vision tokens based on two criteria. The freed space ($|M| - |C|$) accommodates new frames. This process repeats continuously, enabling efficient stream processing under strict memory constraints. When a user query is issued, the model answers using the compressed cache that summarizes visual context from all prior frames. Notably, compression adds only 0.5\% overhead relative to input frames processing time.

InfiniPot-V leverages two token eviction criteria: \textit{Temporal-axis Redundancy (TaR)} and \textit{Value Norm (VaN)} for identifying crucial tokens for compressing KV cache. In the following subsections, we detail each criterion, and finally describe how to effectively combine them.

\subsection{Temporal-axis Redundancy (TaR) Reduction via Patch-wise Similarity}\label{subsec:tar}

Video streams exhibit inherent spatiotemporal redundancy across frames~\cite{h264,shen2024longvu,tao2024dycokedynamiccompressiontokens}.
In this section, we focus on exploiting temporal redundancy, as illustrated in Fig.~\ref{fig:cache_compression}(a) where static patches\footnote{In MLLMs, each vision patch corresponds to a single token, so we use these terms interchangeably.} (e.g., background) persist across frames. For MLLMs processing videos with fixed memory usage, identifying this redundancy in KV caches is crucial. Our analysis in Fig.~\ref{fig:cache_compression}(b) reveals that Key embeddings effectively capture temporal redundancy, exhibiting higher cosine similarity for static patches between adjacent frames compared to Value embeddings, across all layers.

Building on this insight, we propose TaR, a technique that performs a patch-wise comparison of Key embeddings along the temporal axis to detect and reduce redundant tokens. As shown in Fig.~\ref{fig:cache_compression}(c), we introduce a 3D reshaping of Key embeddings to enable direct comparison of corresponding patches across frames. Based on this structured KV cache, the TaR implementation starts with a memory constraint of $|M|$ tokens, processing $f$ consecutive video frames, each containing $p = |M|/f$ vision tokens. To maintain temporal continuity, we designate the $r$ latest frames as \textit{recent frames} and retain them in full. The older \textit{past frames} ($f-r$ frames) are selectively compressed based on their patch-wise similarity to recent frames.

To measure the patch-wise similarity between frames, we divide the current Key embeddings $K \in \mathbb{R}^{H \times (f\times p) \times D}$ into $K_{\text{recent}} \in \mathbb{R}^{H \times r \times p \times D}$ and $K_{\text{past}} \in \mathbb{R}^{H \times (f - r) \times p \times D}$, representing the recent and past frames respectively. For each spatial coordinate $(i, j)$, we compute the $\ell_2$-normalized cosine similarity between recent and past frames of the same patch coordinate:
\begin{equation}\label{eq:tar}
s^{\text{TaR}}(t, i, j) = - \frac{1}{r} \sum_{t'=1}^{r} \text{cos}\Big( K_{\text{past}}^{(t,i,j)}, K_{\text{recent}}^{(t', i, j)} \Big).
\end{equation}
Here, $s^{\text{TaR}}(t, i, j)$ is the importance score of the patch in $t$-th frame at $(i,j)$ coordinate.
The negative sign is applied so that a higher computed score indicates lower redundancy (i.e., the token is more distinctive).
This ensures that tokens with less temporal similarity to recent frames are prioritized.
We then select the least redundant tokens (i.e., higher score) in past frames using the Top-K operator:
\begin{equation}
\mathcal{I_\text{{TaR}}} = \text{TopK}(s^{\text{TaR}}, |C| - |K_{\text{recent}}|),
\end{equation}
where $|C|$ is the target cache compression size and $|K_{\text{recent}}|=rp$ accounts for the recent frame tokens that are always retained.
The compressed key-value pairs are formed by concatenating the selected key frame tokens with all recent frame tokens:
\begin{equation}
\tilde{K}_{\text{TaR}} = \text{Concat}\big(K[:, \mathcal{I_\text{{TaR}}}, :],K_{\text{recent}}\big), \qquad
\tilde{V}_{\text{TaR}} = \text{Concat}\big(V[:, \mathcal{I_\text{{TaR}}}, :],V_{\text{recent}}\big).
\end{equation}
By fully preserving the most recent frames, we maintain complete information on rapidly changing or newly introduced content, while selectively retaining distinctive visual elements from the past.

\subsection{Spatial Semantic Importance Preserving with Value Norm (VaN)}\label{subsec:van}

\begin{wrapfigure}{t}{0.5\columnwidth}
\centerline{\includegraphics[width=0.5\columnwidth]{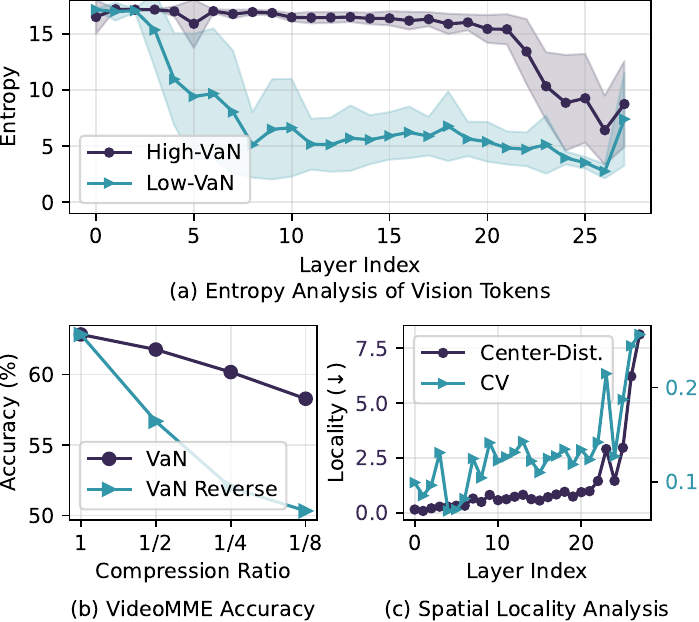}}
\caption{\textbf{Value Norm (VaN) Analysis.} (a) Entropy analysis of vision token representations grouped by their VaN scores. (b) VideoMME performance under varying cache compression ratios using either VaN or reverse-VaN for token selection. (c) Layer-wise locality of VaN, measured by center distance and coefficient of variation (CV); lower values indicate stronger spatial consistency. \texttt{LLaVA-Next-7B} with Video-MME used.}\label{fig:entropy}
\vspace{-.2in}
\end{wrapfigure}

While TaR focuses on reducing temporal redundancy, VaN serves a complementary role: identifying and preserving semantically salient regions within each video frame, independent of the query.
To achieve this, we employ Value embeddings ($V$), which inherently capture semantic information in transformer attention~\cite{NIPS2017_3f5ee243}.
Specifically, we introduce Value Norm (VaN) as a metric for token-level semantic importance: $s^{\text{VaN}} = \|V^{(t,i,j)}\|_2$.

\paragraph{Analysis of Value Norm.}
We hypothesize that tokens with higher VaN contain richer semantic information, making them more valuable for video understanding.
To quantify semantic importance, we project vision token representations from each layer into the vocabulary space~\cite{neo2025towards} and compute the entropy of the resulting word probability distribution, where the higher entropy implies greater informativeness~\cite{farquhar2024detecting,chan2025analyzing}.
As shown in Fig.~\ref{fig:entropy} (a), tokens with higher VaN consistently exhibit higher entropy, confirming their semantic significance.
This advantage translates to improved performance: Fig.~\ref{fig:entropy}~(b) shows that retaining high-VaN tokens achieves substantially higher video understanding accuracy across various compression ratios compared to low-VaN (VaN Reverse) tokens.

\paragraph{Layer-wise Adaptive Pooling.}
An analysis of VaN distributions reveals strong spatial locality patterns in early to middle layers, which gradually diminish in deeper layers as shown in Fig.~\ref{fig:entropy}(c). To measure spatial locality patterns across layers, we employ two methods: (1) compute the average distance between the center point and surrounding points within a $3\times3$ window spanning the VaN values of each frame (center-dist.), and (2) measure the Coefficient of Variance (CV) to quantify dispersion of VaN distributions. Lower values in both metrics—smaller center-dist. and CV—indicate that VaN scores are closely clustered, implying high spatial locality, whereas higher values reflect greater dispersion and lower locality.

As shown in Fig.~\ref{fig:entropy}~(c), both metrics consistently indicate strong locality in early to middle layers, while gradually diminishing in deeper layers. Based on this observation, we design an adaptive spatial pooling mechanism that adjusts the average pooling kernel size per layer. To implement this, we design a mapping function $g$ that assigns kernel sizes in inverse relation to each layer's CV: 
$$ 
\mathrm{PoolSize}(CV_l) = g(CV_l) \quad \text{where} \quad g : \mathbb{R}^+ \rightarrow {1,3,5,7} 
$$ 
This approach assigns larger pooling kernels (e.g., 7) to lower layers with smaller CV values (higher spatial locality), and smaller kernels (e.g., 1, implying no pooling) to upper layers with larger CV values, thus preserving fine-grained details where needed.

For KV cache compression, we select tokens using VaN scores processed through our adaptive pooling mechanism, retaining the Top-$|C|$ tokens with highest pooled VaN values as described in Fig.~\ref{fig:cache_compression}(c): $I_\text{{VaN}} = \text{TopK}(\text{VaN}_{\text{pool}}, |C|)$
\begin{equation}
\tilde{K}_{\text{VaN}} = K[:, \mathcal{I_\text{{VaN}}}, :], \quad \tilde{V}_{\text{VaN}} = V[:, \mathcal{I_\text{{VaN}}}, :].
\end{equation}

\begin{table}[t]
\small
  \centering
  \label{tab:video_qa_comparison}
  \begin{tabular}{l c c c c c c c}
    \toprule
    \textbf{Method}                 & \textbf{Size} & \textbf{\# Frames} & \textbf{Budget} & \textbf{EgoSchema} & \textbf{MLVU} & \textbf{VideoMME} & \textbf{LVB} \\
    Max Duration               &               &                    &             \textbf{$|M|$}    & 3\,min           & 120\,min           & 60\,min          & 60\,min     \\
    \midrule
    GPT4-V*                         & –             & 1fps              & –                       & 55.6               & –            & 60.7              & –           \\
    GPT4-o*                         & –             & 1fps              & –                       & 72.2               & 66.2         & 77.2              & 66.7        \\ \midrule
    LLaVA-OV*                       & 7B           & 32                 & 8K                     & 60.1               & 64.7         & 58.2              & –           \\
    LongVU*                         & 7B           & 1fps              & 8K                     & 67.6               & 65.4         & 60.6              & –           \\
    LongVU*             & 3B           & 1fps              & 8K                     & 59.1               & 55.9         & 51.5              & –           \\
    \midrule
    Qwen-2-VL                       & 7B           & 768                & 50K                    & 65.2              & 65.8        & 63.9             & 58.8       \\
\mscellt Qwen-2-VL + \textbf{Ours}                & 7B           & 768                & 6K                     & 65.6                  & 65.8        & 62.8             & 58.4       \\
    LLaVA-Next                      & 7B           & 128                & 25K                    & 67.6               & 68.7        & 62.8             & 63.5       \\
\mscellt    LLaVA-Next + \textbf{Ours}               & 7B           & 128                & 6K                     & 65.8                  & 65.2        & 61.1             & 60.9       \\
    Qwen-2.5-VL                     & 3B           & 768                & 50K                    & 64.4               & 63.3        & 60.3             & 59.9       \\
\mscellt Qwen-2.5-VL + \textbf{Ours}                & 3B           & 768                & 6K                     & 61.8                   & 62.1        & 59.3             & 56.5       \\
    \bottomrule
  \end{tabular}
  \vspace{2mm}
   \caption{Comparison of various MLLMs accuracy on four Offline Video Understanding (OVU) benchmarks. * denotes the numbers from official paper.}
   \label{tab:open-source}
\end{table}
\begin{table}[t]
\small
  \centering
  \begin{tabular}{lccccccccc}
    \toprule
    \textbf{Compression} & \textbf{Budget} & \multicolumn{3}{c}{\textbf{Video MME}} & \multicolumn{3}{c}{\textbf{MLVU}} & \\
    \cmidrule(lr){3-5} \cmidrule(lr){6-8}
    \textbf{Method} & \textbf{$|M|$} & \textbf{Short} & \textbf{Med} & \textbf{Long} & \textbf{Holistic} & \textbf{Single} & \textbf{Multi} & \textbf{Avg}. \\
    \midrule
    FullKV & 50K & 74.7 & 62.1 & 55.0 & 76.3 & 73.9 & 43.3 & 64.2 \\
    \midrule
    TTC~\cite{tao2024dycokedynamiccompressiontokens} & 3K & 66.8 & 51.2 & 47.9 & 72.1 & 58.8 & 33.2 & 54.8 \\
    (IVC) & 6K & 72.6 & 55.0 & 51.7 & 76.3 & 60.9 & 36.7 & 58.4 \\
    \midrule
    STC~\cite{shen2024longvu} & 3K & 67.9 & 51.0 & 49.3 & 71.5 & 58.6 & 33.9 & 55.0 \\
    (IVC) & 6K & 72.6 & 56.2 & 51.6 & 74.3 & 61.1 & 35.9 & 57.9 \\
    \midrule
    \mscells \textbf{InfiniPot-V} & 3K & 73.9 & 57.8 & 51.8 & 77.7 & 70.4 & 43.2 & 63.1 \\
    \mscellt (CKV) & 6K & 74.1 & 60.8 & 53.4 & 77.2 & 72.3 & 44.8 & 64.3 \\
    \bottomrule
\end{tabular}
\vspace{2mm}
  \caption{Comparison under memory-constrained settings (3, 6K memory-budget) with Input Video Compression (IVC) methods: TTC from DyCoke~\cite{tao2024dycokedynamiccompressiontokens} and STC from LongVU~\cite{shen2024longvu}. \texttt{Qwen-2-VL-7B} used across VIdeoMME and MLVU benchmarks. Results comparing memory-unconstrained IVC methods (without KV cache compression) with InfiniPot-V are provided in Tab.~\ref{tab:full_ivc}.} 
  \label{tab:ic_cache}
\end{table}

\subsection{Design Space Exploration}
\label{subsec:combine}

\textbf{Combining TaR and VaN for Token Selection.}
TaR and VaN capture complementary aspects of spatio-temporal redundancy in streaming video. To integrate them, we prioritize TaR-based selection by first allocating $\alpha |C|$ tokens to TaR, then filling the remaining $(1 - \alpha)|C|$ with VaN-selected tokens. This two-stage selection strategy effectively balances temporal and feature importance. A detailed hyperparameter exploration, including sweeps over $\alpha$ and the size of the recent frame window $r$, is provided in Appendix.~\ref{appn:hyper_parameter}.

\textbf{Comparison with Memory-Constrained Alternatives.}
A natural question is whether IVC or KVC can be adapted for SVU under memory constraints. To explore this, we apply query-agnostic methods such as spatial token compression (STC) and token temporal merging (TTC) from LongVU~\cite{shen2024longvu} and DyCoke~\cite{tao2024dycokedynamiccompressiontokens}. InfiniPot-V outperforms all these baselines by a notable accuracy margin, demonstrating the strength of continual compression over expressive key-value embeddings (details in Tab.~\ref{tab:ic_cache}).

\section{Experiments}\label{sec:exp}

\begin{figure}
\centerline{\includegraphics[width=1\columnwidth]{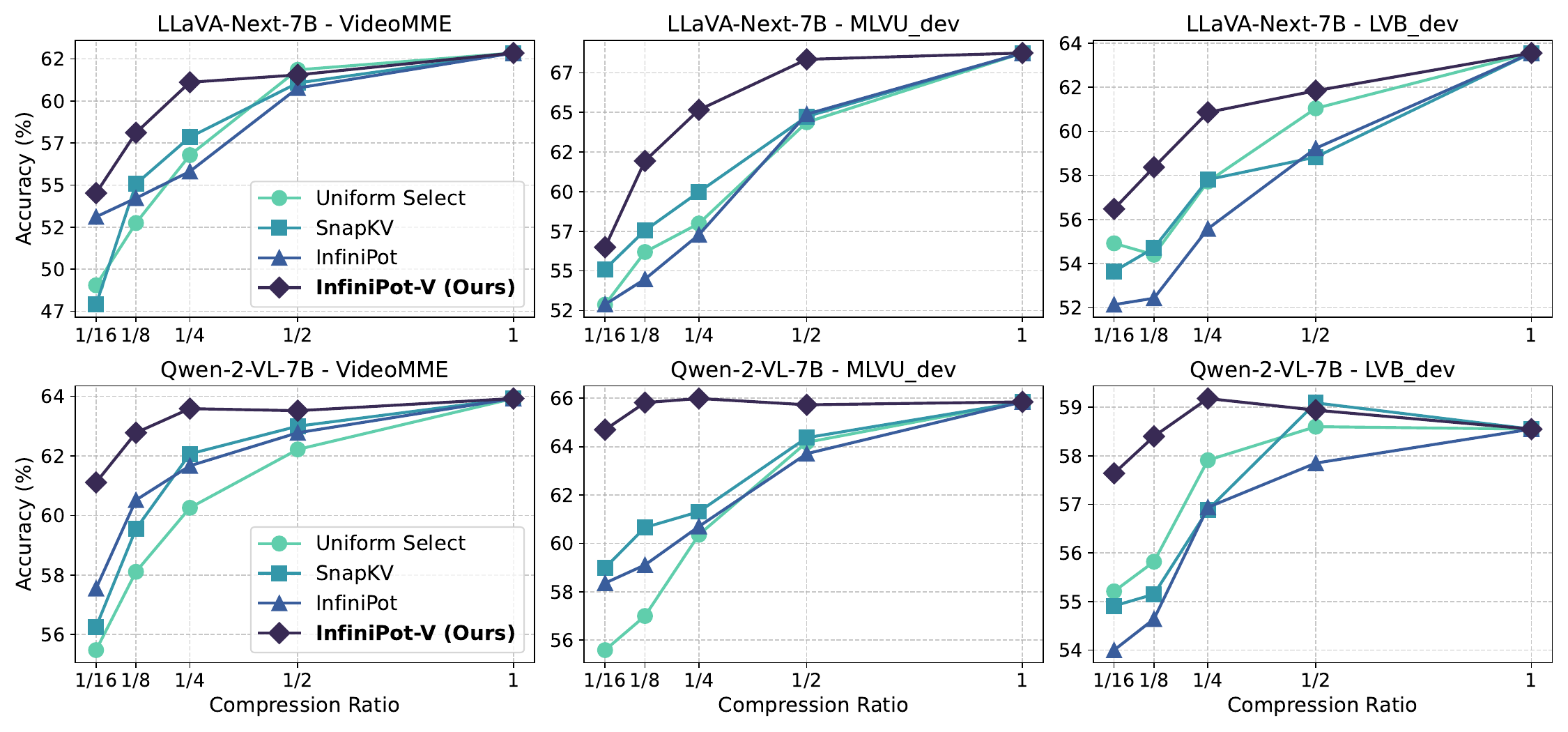}}
\caption{KV cache Compression (KVC) methods evaluation results with offline long video understanding tasks under Continual KV Cache Compression (CKV) framework. Performance across four compression ratios (1/16, 1/8, 1/4, 1/2) for \texttt{LLaVA-Next-7B} (top row) and \texttt{Qwen-2-VL-7B} (bottom row) on VideoMME, MLVU$_\text{dev}$, and LongVideoBench$_\text{dev}$ (LVB$_\text{dev}$) tasks. The full evaluation results are shown in Table~\ref{tab:full_table}.}\label{fig:kvc}
\vspace{-.1in}
\end{figure}

\subsection{Experimental Setup}

\paragraph{Benchmarks.}
We evaluate our InfiniPot-V on both offline video understanding (OVU) and streaming video understanding (SVU) tasks. 
For OVU, we utilize representative multiple-choice based long video understanding benchmarks (ranging from 3 minutes to over 2 hours): VideoMME~\cite{fu2024videomme}, MLVU~\cite{MLVU}, LongVideoBench (LVB)~\cite{wu2024longvideobench}, and EgoSchema~\cite{mangalam2023egoschema}. 

For SVU, we employ RVS-Ego/Movie streaming video QA benchmark~\cite{zhang2024flashvstream}, featuring open-ended questions paired with timestamps, and evaluate the answers using \texttt{GPT-3.5-turbo-0125} following~\cite{zhang2024flashvstream,Shangzhe2025streaming}. 
We further extend multiple-choice based SVU evaluation, OVO-Bench~\cite{li2025ovobenchfarvideollmsrealworld} and StreamingBench~\cite{lin2024streaming}. 
OVO-Bench evaluates temporal reasoning under three scenarios—backward tracing, real-time visual perception, and forward active responding—while StreamingBench evaluates real-time visual comprehension of streaming videos.

\paragraph{Models.}
We apply our method on four state-of-the-art MLLMs capable of long-video understanding: Qwen-2-VL~\cite{Qwen2VL}, Qwen-2.5-VL~\cite{qwen2.5}, LLaVA-OV-7B~\cite{li2024llava}, and LLaVA-Next-Video~\cite{zhang2024llavanext-video}. Details on input video sampling settings and benchmark details are provided in Appendix.~\ref{appn:exp}.

\subsection{Evaluatal Results}
\paragraph{Offline Video Understanding. } 
To assess the absolute compression capability of our method, we evaluate InfiniPot-V with both commercial MLLMs (GPT-4V~\cite{openai2023gpt4v}, GPT-4o~\cite{openai2023gpt4o}) and state-of-the-art public models designed for offline video understanding, including LLaVA-OV~\cite{li2024llava}, and LongVU~\cite{shen2024longvu}. Unlike these specialized, fully trained models, InfiniPot-V is a training-free, plug-in framework compatible with MLLMs of various scales, enabling high performance under fixed memory budgets.
As shown in Tab.~\ref{tab:open-source}, InfiniPot-V reduces memory usage to just 25\% (6K tokens) for LLaVA-Next (originally 25K tokens) and 12.5\% for Qwen-VL series (6K vs. 50K), with minimal performance loss. Notably, it achieves comparable or better accuracy than LongVU at the 7B scale and significantly outperforms it at 3B, demonstrating both efficiency and scalability.

\paragraph{Comparison with IVC under Memory Constraints.}
To evaluate recent query-agnostic IVC methods under memory-constrained CKV, we adopt a unified setup on VideoMME and MLVU: token temporal merging (TTC) from DyCoke~\cite{tao2024dycokedynamiccompressiontokens} and spatial token compression (STC) from LongVU~\cite{shen2024longvu} are applied to compress vision tokens to fit the target memory budget $|M|$, while KV cache is managed using a sliding window attention (SWA)~\cite{Beltagy2020Longformer}. When operated under such constraints, these IVC methods suffer from notable accuracy degradation. In contrast, InfiniPot-V performs KV cache compression using TaR and VaN, leveraging expressive key-value representations to achieve superior average accuracy under a 6K memory budget—corresponding to an 88\% lossless compression rate.

\paragraph{Comparison with KVC under Memory Constraints.}
Fig.\ref{fig:kvc} evaluates KVC methods within our CKV framework under constrained memory across offline video understanding tasks. Compression ratios (1/16, 1/8, 1/4, 1/2) are defined based on each model's maximum frame capacity (e.g., 128 frames for LLaVA-Next, 768 for Qwen-2-VL). Our InfiniPot-V consistently outperforms all baseline methods (Uniform Select, SnapKV, InfiniPot) across all tasks for both LLaVA-Next-7B and Qwen-2-VL-7B, demonstrating superior video understanding performance. Under CKV constraints—where actual query access is not available—query-dependent methods like SnapKV\cite{li2024snapkv} degrade significantly. In contrast, InfiniPot-V maintains strong accuracy even at high compression ratios (e.g., 1/16), thanks to its query-agnostic selection via TaR and VaN.

\begin{table}[t]
    \small
  \centering
  \begin{tabular}{l c c c c c c c c c}
    \toprule
    & \multicolumn{2}{c}{\textbf{RVS-Ego}} & \multicolumn{2}{c}{\textbf{RVS-Movie}} & \textbf{Execution Time }
    & \multicolumn{2}{c}{\textbf{Total Memory Usage}} \\ 
    \cmidrule(lr){2-3} \cmidrule(lr){4-5} \cmidrule(lr){6-6} \cmidrule(lr){7-8}
       \textbf{LLaVA-OV-7B} & \textbf{Acc} & \textbf{Score} & \textbf{Acc} & \textbf{Score} & \textbf{Video Enc. (msec/Frame)} 
      & \textbf{GPU} & \textbf{CPU} \\
    \midrule
    ReKV                   & 60.1 & 3.9 & 53.4 & 3.8  &  285.7  & 37.5 GB & + 18.8GB/h \\ \midrule
    ReKV w/o off.       & 55.8 & 3.3 & 50.8 & 3.4  & 74.6  & 27.2 GB     & 0          \\
    InfiniPot-V                   & \textbf{57.9} & \textbf{3.5} & \textbf{51.4} & \textbf{3.5}  & 76.3  & 27.8 GB     & 0          \\
    \bottomrule
  \end{tabular}
  \vspace{2mm}
  \caption{Streaming benchmark comparison to offloading-based KV cache control method. (ReKV) Video Enc. shows execution time per frame, GPU indicates peak memory usage, and CPU denotes the size of video KV-Cache offloaded to CPU per hour. Results based on an 1-hour video processed with a 0.5 fps sampling rate in streaming mode. \texttt{LLaVA-OV-7B is used.}}
  \vspace{-.2in}
  \label{tab:rvs_ego_performance}
\end{table}
\begin{table}[t]
\small
\centering
\begin{tabular}{l c c c c c c c}
\toprule
\textbf{Qwen-2.5-VL} & \textbf{Scale}
& \textbf{OVO-BW} & \textbf{OVO-Real} & \textbf{OVO-FW} 
& \textbf{OVO-Avg} & \textbf{StreamingBench} \\
\midrule
Uniform Select & 7B & 44.5 & 61.1 & 47.8 & 51.7 & 75.2 \\
InfiniPot-V & 7B & 47.6 & 65.9 & 47.9 & \textbf{53.6} & \textbf{76.4} \\
\bottomrule
\end{tabular}
\vspace{.1in}
\caption{Comparison on OVO-Bench (BW: Backward, Real: Realtime, FW: Forward) and StreamingBench (Real-time visual understanding) using \texttt{Qwen-2.5-VL-7B} under 4K memory budget.}
\label{tab:ovo_results}
\vspace{-.2in}
\end{table}

\paragraph{Streaming Video Understanding.}
We first evaluate InfiniPot-V on streaming video understanding (SVU) using two  StreamingVQA benchmarks, RVS-Ego and RVS-Movie, with LLaVA-OV-7B.
As a baseline, we compare against ReKV~\cite{Shangzhe2025streaming}, a state-of-the-art SVU method, under two configurations: 
(1) CPU–GPU with CPU offloading, which allows spilling KV cache to CPU memory, and 
(2) CPU–GPU without offloading, simulating shared-memory devices where CPU memory is unavailable or pre-occupied~\cite{Karumbunathan2022JetsonAGXOrin}. 
Tab.~\ref{tab:rvs_ego_performance} reports SVU accuracy, compression time, and memory usage. 
While CPU offloading enables ReKV to retain the full KV cache, it incurs substantial transfer overhead; without offloading, ReKV suffers from severe accuracy degradation due to limited local memory. 
In contrast, InfiniPot-V operates entirely within GPU memory, eliminating offloading latency and achieving higher accuracy, making it a practical solution for memory-constrained systems.

To further evaluate accuracy in streaming scenarios, we extend the analysis to two additional benchmarks: OVO-Bench~\cite{li2025ovobenchfarvideollmsrealworld} and StreamingBench~\cite{lin2024streaming}, as summarized in Tab.~\ref{tab:ovo_results}. 
Compared to the uniform token selection baseline, InfiniPot-V demonstrates consistent gains across all metrics, notably improving recall of past visual information on the OVO-BW (backward) task (44.5 $\rightarrow$ 47.6) and achieving higher real-time understanding accuracy on the OVO-Real (Realtime) (61.1 $\rightarrow$ 65.9) and StreamingBench score (75.2 $\rightarrow$ 76.4), demonstrating the effectiveness of our TaR–VaN compression in eliminating temporal redundancy while preserving spatially salient semantics, showing its capability under challenging SVU settings.

\subsection{Ablation Study}
\begin{table}[t]
\small
  \centering
  \begin{tabular}{l c c c c c c c c}
    \toprule
    \textbf{MLVU}$_\text{dev} $& \multicolumn{2}{c}{\textbf{Holistic Reasoning}} 
    & \multicolumn{3}{c}{\textbf{Single Detail}} 
    & \multicolumn{2}{c}{\textbf{Multi Detail}} \\
    \cmidrule(lr){2-3} \cmidrule(lr){4-6} \cmidrule(lr){7-8}
    \textbf{Ablation Study} 
      & \textbf{Topic}  & \textbf{Anomaly} 
      & \textbf{Plot}   & \textbf{Needle} 
      & \textbf{Ego}    & \textbf{AO} 
      & \textbf{AC}     & \textbf{Avg} \\
    \midrule
    Full KV           & 85.2 & 67.5 & 72.7 & 83.9 & 65.1 & 54.1 & 32.5 & 65.9 \\ \midrule
    Uniform Select    & 83.7 & 66.5 & 67.9 & 76.1 & 58.5 & 51.0 & 27.2 & 61.5 \\ \midrule
    TaR Reverse       & 79.0 & 64.5 & 56.9 & 65.6 & 55.1 & 45.2 & 21.8 & 55.5 \\
    TaR Frame         & 82.9 & 66.0 & 67.0 & 78.9 & 63.6 & 51.0 & 31.1 & 62.9 \\
    TaR               & 85.9 & 66.5 & 71.8 & 78.0 & 62.2 & 51.7 & 35.4 & \textbf{64.5} \\ \midrule
    VaN Reverse       & 78.3 & 66.5 & 56.2 & 66.8 & 53.4 & 46.3 & 17.5 & 55.0 \\ 
    VaN               & 84.4 & 68.0 & 68.6 & 76.6 & 61.9 & 52.5 & 29.1 & 63.0  \\ 
    VaN + Pool        & 85.2 & 68.0 & 71.4 & 77.5 & 63.1 & 52.1 & 31.5 & \textbf{64.1} \\ \midrule
    TaR + VaN + Pool   & 86.3 & 68.0 & 72.7 & 80.3 & 63.9 & 54.1 & 35.4 & \textbf{65.8} \\
    \bottomrule
    \vspace{1mm}
  \end{tabular}
  \caption{Ablation study of TaR, VaN, and their combination. Experiments conducted on MLVU using \texttt{Qwen-2-VL-7B} with a 6K memory budget.}
    \vspace{-.2in}
\label{tab:ablate}
\end{table}

Tab.~\ref{tab:ablate} validates our design decisions for TaR and VaN. Reversed strategies (TaR Reverse and VaN Reverse) significantly degrade performance by discarding distinctive or semantically important tokens. Within TaR, patch-wise similarity proves more effective than frame-level similarity (64.5 vs. 62.9). VaN alone surpasses the baseline, and its performance improves further with adaptive pooling (64.1 vs. 63.0). Combining TaR and VaN yields the highest accuracy, significantly outperforming the baseline. Additional integration explorations are discussed in Appendix.~\ref{appn:hyper_parameter}.

\begin{table}[t]
\small
\centering
\begin{tabular}{l c c c c c c}
\toprule
\textbf{Method} & \textbf{Length} & \textbf{Context} &
\textbf{VRAM} & \textbf{FPS} & \textbf{Throughput} & \textbf{Power} \\
 & \textbf{(sec)} & \textbf{Length} &
\textbf{(GB) ↓} & \textbf{(frame/sec) ↑} & \textbf{(tok/sec) ↑} & \textbf{(J) ↓} \\
\midrule
Full KV & 100 & 18K & 13.8 & 6.2 & 5.0 & 1.6 \\
InfiniPot-V    &     &     & 9.2 (1.5$\times$) & 6.4 (1.0$\times$) & 9.2 (1.8$\times$) & 1.2 (1.4$\times$) \\ \midrule
Full KV & 300 & 54K & 26.6 & 5.0 & 2.0 & 4.7 \\
InfiniPot-V    &     &     & 10.1 (2.6$\times$) & 6.4 (1.3$\times$) & 9.2 (4.6$\times$) & 2.4 (2.0$\times$) \\ \midrule
Full KV & 500 & 90K & 39.0 & 4.1 & 1.2 & 7.5 \\
InfiniPot-V    &     &     & 10.7 (3.6$\times$) & 6.3 (1.5$\times$) & 9.1 (7.3$\times$) & 3.7 (2.0$\times$) \\ \midrule
Full KV & 600 & 108K & OOM & OOM & OOM & OOM \\
InfiniPot-V    &      &      & 11.3 & 6.4 & 9.2 & 4.3 \\
\bottomrule
\end{tabular}
\vspace{.1in}
\caption{Streaming video processing performance results on NVIDIA Jetson AGX Orin. Streaming 10-minute video samples (FPS 0.2–1) processed with Qwen-2.5-VL-3B, measuring memory, FPS (frames processed per second), throughput, and power consumption.}
\label{tab:jetson}
\vspace{-.2in}
\end{table}

\subsection{Edge Device Deployment Results}
\label{sec:deployment}

We evaluate our CKV framework—combining continual KV compression with TaR and VaN scoring—on the NVIDIA Jetson AGX Orin~\cite{Karumbunathan2022JetsonAGXOrin} using \texttt{Qwen-2.5-VL-3B} and 10-minute StreamingBench~\cite{lin2024streaming} videos (0.2–1 FPS). As shown in Tab.~\ref{tab:jetson}, our method achieves consistent efficiency across all metrics. Peak memory remains nearly constant (9.2–10.7\,GB) while Full KV grows linearly (13.8→39.0\,GB), yielding a 3.6$\times$ reduction at 500 seconds streaming video. Prefill speed, measured in FPS (frames processed per second during prefill), stays stable at 6.3–6.4\,FPS, whereas Full KV drops to 4.1\,FPS. Generation throughput also increases up to 7.3$\times$ (1.2→9.1\,tok/sec) under a fixed memory budget. Power usage scales proportionally with compute, improving energy efficiency by nearly 2$\times$ (7.5→3.7\,J). Notably, CKV enables continuous inference on 600-second streams where Full KV fails with OOM, confirming its practicality for real-time multimodal reasoning on memory-constrained edge devices.

\section{Conclusion}\label{sec:conclustion}

In this paper, we proposed InfiniPot-V, a training-free KV cache control framework for streaming video processing in memory-constrained environments.
Built around practical constraints—unavailable queries and strict memory budgets during compression—InfiniPot-V employs two novel token eviction criteria, TaR and VaN, achieving significant improvements in long video understanding under streaming scenarios.

\bibliographystyle{plain}
\bibliography{refs}

\renewcommand{\thefigure}{A\arabic{figure}}
\renewcommand{\thetable}{A\arabic{table}}
\appendix

\clearpage
\setcounter{figure}{0}
\setcounter{equation}{0}
\setcounter{table}{0}
\setcounter{section}{0}
\appendix

\newpage
\begin{algorithm}[t]
\footnotesize
\caption{InfiniPot-V Algorithm}
\label{alg:infinipot_v}
\begin{algorithmic}[1]
\Require Video stream ${V}$, memory constraint $|M|$, target KV cache size $|C|$, recent frame count $r$, CV thresholds $\{\tau_1, \tau_2, \tau_3\}$, TaR ratio $\alpha \in [0,1]$, $f$ frames corresponding to $|M|$ tokens, vision token number per single frame $p=|M|/f$ 
\Ensure Compressed KV cache $\{\tilde{K}_l, \tilde{V}_l\}_{l=1}^L$
\State Let $C_{\text{TaR}} = \alpha|C|$ be the TaR selection budget
\State Let $C_{\text{VaN}} = (1-\alpha)|C|$ be the VaN selection budget
\State Initialize empty KV cache for each layer $l \in \{1,\ldots,L\}$
\While{processing video stream ${V}$}
    \State Accumulate KV embeddings until reaching $|M|$ 
    \For{each layer $l$}
        \State \textcolor{teal}{// Temporal-axis Redundancy (TaR)}
        \State Reshape $K_l$ into $K_{\text{recent},l} \in \mathbb{R}^{H \times r \times p \times D}$ and $K_{\text{past},l} \in \mathbb{R}^{H \times (f-r) \times p \times D}$
        \For{each patch $(t,i,j)$ in past frames}
            \State $s(t,i,j) = -\frac{1}{r}\sum_{t'=1}^{r}\text{cos}(K_{\text{past},l}^{(t,i,j)}, K_{\text{recent},l}^{(t',i,j)})$
        \EndFor
        \State $\mathcal{I}_l \gets \text{TopK}(S_l, C_{\text{TaR}})$ \Comment{Select least redundant tokens}
        
        \State \textcolor{teal}{// Value Norm (VaN) with Adaptive Pooling}
        \State $\text{VaN}_l \gets \|V_l\|_2$
        \State \textcolor{teal}{// Compute CV for adaptive pooling}
        \State $\mu_l \gets \text{mean}(\text{VaN}_l)$
        \State $\sigma_l \gets \text{std}(\text{VaN}_l)$
        \State $\text{CV}_l \gets \sigma_l / \mu_l$
        \State \textcolor{teal}{// Determine pooling size using mapping function $g$}
\State $\text{pool\_size}_l \gets g(\text{CV}_l)$ \Comment{Using thresholds $\{\tau_1, \tau_2, \tau_3\}$}
\State \textbf{where} $g(\text{CV}) = 
\begin{cases}
    7, & \text{if}\ \text{CV} < \tau_1 \\
    5, & \text{if}\ \tau_1 \leq \text{CV} < \tau_2 \\
    3, & \text{if}\ \tau_2 \leq \text{CV} < \tau_3 \\
    1, & \text{if}\ \text{CV} \geq \tau_3
\end{cases}$
        \State $\text{VaN}_{pool,l} \gets \text{AveragePool2d}(\text{VaN}_l, \text{pool\_size}_l)$
        
        \State \textcolor{teal}{// Combine TaR and VaN by prioritizing TaR-selected tokens}
        \State $\text{VaN}_{pool,l}[\mathcal{I}_l] \gets \max(\text{VaN}_{pool,l})$ \Comment{Prioritize TaR tokens}
        \State $\mathcal{J}_l \gets \text{TopK}(\text{VaN}_{pool,l}, |C|)$ \Comment{Final token selection}
        \State $\tilde{K}_l \gets K_l[:, \mathcal{J}_l, :]$, $\tilde{V}_l \gets V_l[:, \mathcal{J}_l, :]$ \Comment {Update layer KV cache with compressed KV cache}
    \EndFor
\EndWhile
\end{algorithmic}
\end{algorithm}

\begin{table}[t]
\centering
\small
\begin{tabular}{lcccccc}
\toprule
VideoMME  &\multicolumn{6}{c}{$|M| = \alpha |\text{TaR}| + (1-\alpha) |\text{VaN}|$, $|M| = 6\text{K}$} \\ \midrule
Qwen-2-VL-7B & $\alpha=0$ (VaN) & $\alpha=0.2$ & $\alpha=0.4$ & $\alpha=0.6$ & $\alpha=0.8$ & $\alpha=1$ (TaR) \\
\cmidrule(lr){1-7}
Short (-3 min) & 74.4 & 74.3 & 74.1 & 74.9 & 74.4 & 73.8 \\
Medium (3 - 30 min)  & 59.9 & 59.3 & 61.3 & 61.4 & 61.2 & 58.2 \\
Long (30 - 120 min)  & 51.9 & 51.0 & 52.4 & 53.1 & 53.4 & 53.2 \\ \cmidrule(lr){1-7}
Average  & 62.1 & 61.6 & 62.6 & \textbf{63.1} & \underline{63.0} & 61.7 \\
\midrule
LLaVA-Next-7B & $\alpha=0$ (VaN) & $\alpha=0.2$ & $\alpha=0.4$ & $\alpha=0.6$ & $\alpha=0.8$ & $\alpha=1$ (TaR) \\
\cmidrule(lr){1-7}
Short (-3 min)   & 69.8 & 69.8 & 72.0 & 71.1 & 71.9 & 68.8 \\
Medium (3 - 30 min)     & 59.3 & 59.3 & 59.2 & 58.7 & 57.7 & 57.3 \\
Long (30 - 120 min)    & 52.1 & 52.1 & 52.7 & 52.0 & 51.4 & 51.0 \\ \cmidrule(lr){1-7}
Average    & 60.4 & 60.4 & \textbf{61.3} & \underline{60.6} & 60.3 & 59.0 \\
\bottomrule
\end{tabular}
\vspace{2mm}
\caption{\textbf{TaR and VaN Combination Ratio}: Sweep over combination ratio $\alpha$ in TaR and VaN combination under a 6K memory budget ($|M|$) on VideoMME. $\alpha{=}0$ and $\alpha{=}1$ correspond to VaN-only and TaR-only, respectively. The best-performing configurations are shown in \textbf{bold}, while the second-best results are \underline{underlined}.}
\label{tab:alpha_sweep}
\vspace{-.15in}
\end{table}

\begin{table}[t]
\centering
\small
\begin{tabular}{lcccccccc}
\toprule
Qwen-2-VL-7B & \multicolumn{4}{c}{MLVU}            & \multicolumn{4}{c}{VideoMME}      \\
 \cmidrule(lr){1-1} \cmidrule(lr){2-5} \cmidrule(lr){6-9}
 $|M| = 6\text{K}$  & Holistic & Single & Multi & Avg. & Short & Medium & Long & Avg. \\
\midrule
$r=f\times0.125$        & 77.6 & 66.2 & 43.9 & \underline{63.1} & 68.7 & 57.3 & 51.0 & \textbf{59.0} \\
$r=f\times0.25$         & 77.8 & 67.1 & 43.5 & \textbf{63.4} & 68.0 & 57.2 & 51.1 & \underline{58.8} \\
$r=f\times0.50$          & 78.6 & 64.6 & 39.0 & 61.3 & 65.1 & 56.7 & 49.7 & 57.2 \\
\midrule
$|M|/|C|=0.75$   & 78.2 & 71.7 & 44.6 & \textbf{65.8} & 74.4 & 59.9 & 51.9 & \underline{62.1} \\
$|M|/|C|=0.50$   & 76.8 & 68.2 & 42.3 & \underline{63.3} & 74.1 & 60.3 & 52.9 & \textbf{62.4} \\
$|M|/|C|=0.25$   & 73.2 & 65.9 & 39.1 & 60.3 & 73.6 & 56.9 & 50.4 & 60.3 \\
\bottomrule
\end{tabular}
\vspace{2mm}
\caption{\textbf{Recent Frame and Compression Ratio Exploration}: \textbf{Top}: Sweep over recent frame numbers $r$ determined by multiplying various ratios ($0.125, 0.25, 0.5$) to $f$ (frame number corresponding to memory budget $|M|$) in TaR. \textbf{Bottom}: Performance under varying compression ratios $|M|/|C|$ across MLVU and VideoMME with \texttt{Qwen-2-VL-7B}. TaR performs best with $r\leq0.25$ and compression ratio $\geq0.5$. The highest values are shown in \textbf{bold}, with the second-highest \underline{underlined}.}
\label{tab:recent_compress_ratio}
\vspace{-.2in}
\end{table}

\section{InfiniPot-V Algorithm and Configuration}\label{appn:algorithm}

\subsection{Algorithm Description} Algorithm~\ref{alg:infinipot_v} presents the complete process of InfiniPot-V's cache control framework along with its compression formulation. InfiniPot-V processes video streams by continuously pre-filling and compressing the KV cache using two token selection strategies: Temporal-axis Redundancy (TaR) and Value Norm (VaN). For TaR, the algorithm splits video frames into recent frames (the latest $r$ frames) and past frames, then computes cosine similarities between corresponding patches to identify and remove redundant visual tokens. (Line 10)

For spatial semantic importance token selection, a layer-wise adaptive pooling mechanism based on VaN is employed. The pooling size is dynamically determined by the Coefficient of Variation (CV) of the VaN, (Line 18) where a higher CV indicates a sparser or more distinct feature distribution. Pre-computed model-specific CV thresholds $\{\tau_1, \tau_2, \tau_3\}$ determine pooling sizes from the set \{1,3,5,7\}, selecting larger windows for uniform (low CV) VaN distributions and smaller ones for sparse (high CV) VaN distributions (Line 21).

To integrate both criteria, TaR-selected tokens are prioritized by assigning them the maximum VaN score before the final token selection. Specifically, by setting $\text{VaN}_{pool,l}[\mathcal{I}_l] = \max(\text{VaN}_{pool,l})$ (Line 24) and then applying a TopK selection, the algorithm ensures that temporally distinctive tokens are preserved while allowing VaN to select additional tokens based on semantic importance.

\subsection{Hyper-Parameter Exploration}\label{appn:hyper_parameter} InfiniPot-V involves three main hyper-parameters: the TaR and VaN budget allocation ratio $\alpha$, the number of recent frames $r$ used in TaR, and the target compression size $C$ applied at each continual KV cache compression step. This section presents comparative experiments exploring each hyper-parameter.

\paragraph{TaR and VaN Budget Ratio ($\alpha$)} We compare the accuracy of offline video understanding (OVU) task across different values of $\alpha$, which determines the budget allocation between TaR and VaN under a fixed memory budget ($|M|=6\text{K}$), for both Qwen-2-VL-7B and LLaVA-Next-7B models. As shown in Tab.~\ref{tab:alpha_sweep}, performance peaks when $\alpha$ is between 0.4 and 0.6, outperforming the use of either VaN-only ($\alpha=0$) or TaR-only ($\alpha=1$). This confirms the effectiveness of our approach, which jointly considers both spatial and temporal dimensions for KV cache compression.

\paragraph{Recent Frames ($r$) and Compression Ratio ($|M|/|C|$)} Tab.~\ref{tab:recent_compress_ratio} presents exploration experiments for two key hyperparameters: the recent frames number $r$, which determines the proportion of recent frames within the memory budget in TaR, and the compression ratio $|M|/|C|$, which defines what proportion of the compression size $|C|$ to maintain relative to the memory budget $|M|$ in continual KV cache compression (CKV). 

For the recent frame number $r$ (Tab.~\ref{tab:recent_compress_ratio}, Top), we observe optimal performance on both MLVU and VideoMME benchmarks when $r \leq 0.25f$. Setting $r=0.5f$ results in an excessive number of frames being designated as the latest frames for temporal redundancy measurement, which limits the effectiveness of redundancy reduction. This limitation is reflected in the decreased performance metrics. (61.3 vs 63.4 in MLVU and 57.2 vs 59.0 in VideoMME) Note that the $r$ sweep experiments are conducted using TaR-only settings ($\alpha=1$).

For the compression ratio ($|M|/|C|$), we conduct comparative experiments across three ratios (0.75, 0.50, and 0.25). As shown in Tab.~\ref{tab:recent_compress_ratio} Bottom, an excessive compression ratio such as 0.25 in CKV results in noticeable performance degradation. These findings confirm that a ratio of 0.5 or higher represents an appropriate configuration for CKV.

Based on these explorations, we standardize the hyperparameter values at $\alpha=0.5$, $r=0.125$, and $|M|/|C|=0.75$ for all main experimental results when evaluating InfiniPot-V.

\section{Experimental Setting Details}\label{appn:exp}

\subsection{MLLMs Video Sampling Details.} For all benchmarks, we employ a consistent uniform frame sampling strategy to ensure maximized long video understanding performance across all settings. For Qwen-2-VL~\cite{Qwen2VL}, which supports dynamic image resizing based on the number of frames, we use the hyper-parameter configuration reported to yield the best performance in their original work: \texttt{FPS\_MAX\_FRAMES} = 768, \texttt{VIDEO\_MIN\_PIXEL} = ${128 \times 28 \times 28}$ and \texttt{VIDEO\_MAX\_PIXEL} = ${768 \times 28 \times 28}$. Although theoretically larger token budgets could be set, we adopt this configuration to match the optimal context length of 50K as reported in the original paper~\cite{Qwen2VL}, on top of which we applied KV cache compression. For LLaVA, we set the number of sampled frames to 128 to ensure it remained within the model’s trained context length (<32K). With this video sampling configuration, Qwen-2-VL~\cite{Qwen2VL} uses 384 frames with 130 tokens per frame, resulting in a total context length of 49,920 tokens, while LLaVA-Next~\cite{zhang2024llavanext-video} and LLaVA-OV~\cite{li2024llava} use 128 frames with 196 tokens per frame, yielding a total of 25,088 for offline long video inputs.

\subsection{Long Video Understanding Benchmark Details} 
\paragraph{Offline Video Understanding(OVU)} We evaluate our method on four multiple-choice based offline video question answering benchmarks: Video-MME~\cite{fu2024videomme}, MLVU~\cite{MLVU}, EgoSchema~\cite{mangalam2023egoschema}, and LongVideoBench~\cite{wu2024longvideobench}. For MLVU and EgoSchema, we use the development sets for evaluation.

For Video-MME, we report results without subtitles version. This is because prepending subtitles for all video frames as a single context block directly before the question represents an unrealistic setting that is incompatible with streaming scenarios, where subtitles are typically unavailable during real-time video processing and would not be accessible as complete context in advance.

\paragraph{Streaming Video Understanding (SVU)} For SVU evaluation, we use two benchmarks: RVS-Ego and EVS-Movie~\cite{zhang2024flashvstream}. RVS-Ego is constructed from 10 videos from the Ego4D~\cite{Grauman_2022_CVPR} dataset, while RVS-Movie uses 22 long videos from MovieNet~\cite{huang2020movienet}. Each benchmark consists of a QA set containing open-ended generation questions and their corresponding timestamps indicating when each question should be presented during video streaming.

The evaluation process works as follows: during CKV processing, when the video stream reaches the timestamp of a given question sample, we present the question and generate an answer based on the compressed KV cache accumulated up to that point. The generated answers are then compared against ground-truth answers using \texttt{GPT-3.5-turbo-0125} to produce accuracy and score metrics.

\subsection{Baseline Settings}
\paragraph{Input Video Compression (IVC) Details.}
For the comparison with Input Video Compression (IVC) methods in Tab.~\ref{tab:ic_cache} and ~\ref{tab:full_ivc}, we implement LongVU~\cite{shen2024longvu} and DyCoke~\cite{tao2024dycokedynamiccompressiontokens} as follows: For \textbf{LongVU}, we apply Spatial Token Compression (STC) every 8 frames as specified in the original paper. STC compresses vision token embeddings by identifying temporally redundant patches using cosine similarity between patches. We adjust the similarity threshold to control the compression rate while maintaining the original methodology. For \textbf{DyCoke}, we implement Token Temporal Merging (TTM) which, similar to LongVU, compresses vision encoder output features. TTM calculates cosine similarity between patches in adjacent frames to eliminate redundant patch embeddings. Following the original paper, we process compression every 4 frames and adjust the similarity threshold to control compression rates.

For fair comparison in the Continual KV cache compression (CKV) framework in Tab.~\ref{tab:ic_cache}, we adapt both methods to work within memory constraint $|M|$. Specifically, we compress each input video stream to size $|C|$ and implement sliding window attention~\cite{Beltagy2020Longformer} to evict older KV cache entries once the cache size reaches the predefined memory limit ($|M|$). This adaptation ensures all methods operate under identical memory constraints for a fair comparison with InfiniPot-V. For benchmark comparing InfiniPot-V with IVC methods that use full vision encoding without cache compression, see Tab.~\ref{tab:full_ivc}.

\paragraph{KV Cache Compression (KVC) Details.} In Fig.~\ref{fig:cache_compression}, we compare three KV cache compression methods within Continual KV cache compression (CKV). First, \textbf{Uniform Select}, inspired by uniform video sampling approaches, selects frames at regular intervals and retains all KV cache tokens corresponding to those frames. For \textbf{SnapKV}~\cite{li2024snapkv}, we follow the original method configuration under the CKV process, using the last 32 tokens of the given budget $|M|$ tokens as an observation window ($w$) to calculate attention scores for token selection (see Eq.~\ref{eq:importance-future} in Appendix.~\ref{appn:prelim}). Additionally, we apply 1D pooling with a kernel size of 7 to these scores, as done in the original implementation\footnote{https://github.com/FasterDecoding/SnapKV}. For \textbf{InfiniPot}~\cite{kim-etal-2024-infinipot}, we design a proxy prompt for video compression: "Provide a detailed description of this video." This prompt is utilized in the CaP method to generate attention scores and apply KV cache compression. Detailed experimental results are provided in Tab.~\ref{tab:full_table}.

\paragraph{FastV Hyper-Parameter Settings.} To provide additional performance comparison with compression methods specialized for MLLMs, we also include performance comparisons with FastV~\cite{chen2024image} in Tab.\ref{tab:full_table}. FastV requires two hyper-parameters, $L$ and $R$, which specify the layer where token pruning begins and the percentage of tokens to prune. For a fair comparison, we adjust the $R$ of FastV to ensure that the total number of KV cache entries across layers matches the total entry count of other baselines that maintain the same number of KV-cache entries across each layer.  Specifically, for Qwen-2-VL, the $(L, R)$ pairs corresponding to memory budgets of 3K, and 6K are set to (2, 2.8\%) and (2, 5.8\%) respectively.

\subsection{Positional Encoding Details.} MLLM backbone LLMs utilize positional encoding to differentiate vision token positions. LLaVA-Next~\cite{zhang2024llavanext-video} and LLaVA-OV~\cite{li2024llava} use standard 1D RoPE~\cite{su2023roformerenhancedtransformerrotary}, while Qwen-2-VL~\cite{Qwen2VL} employs 3D RoPE for multimodal encoding. For Offline Video Understanding(OVU), we apply KV cache compression after positional encoding (i.e., Post RoPE). However, Streaming Video Understanding(SVU) presents a challenge: continuous video stream processing can exceed the model's maximum positional range. For example, in LLaVA models with 196 tokens per frame, streaming more than 6 minutes of video at 0.5 FPS exceeds the 32K context window (note that RVS-Ego and RVS-Movie average over 60 minutes). 

To address this, we adopt strategies from InfiniPot~\cite{kim-etal-2024-infinipot} and ReKV~\cite{Shangzhe2025streaming}, re-ordering positional indices to fit within the memory budget $|M|$ at each CKV step. Specifically, we cache the pre-positional encoded KV's hidden states and re-assign positional indices during decoding, ensuring they never exceed $|M|$ position regardless of video length. While this enables SVU for arbitrarily long videos, it discards the original positional information of vision tokens. In particular, additional handling is required for Qwen-2-VL’s 3D RoPE. Developing methods that preserve the original spatial and temporal position encoding while supporting streaming video lengths beyond the model’s positional capacity remains an open direction for future work.


\begin{figure*}[t]
\centerline{\includegraphics[width=1\textwidth]{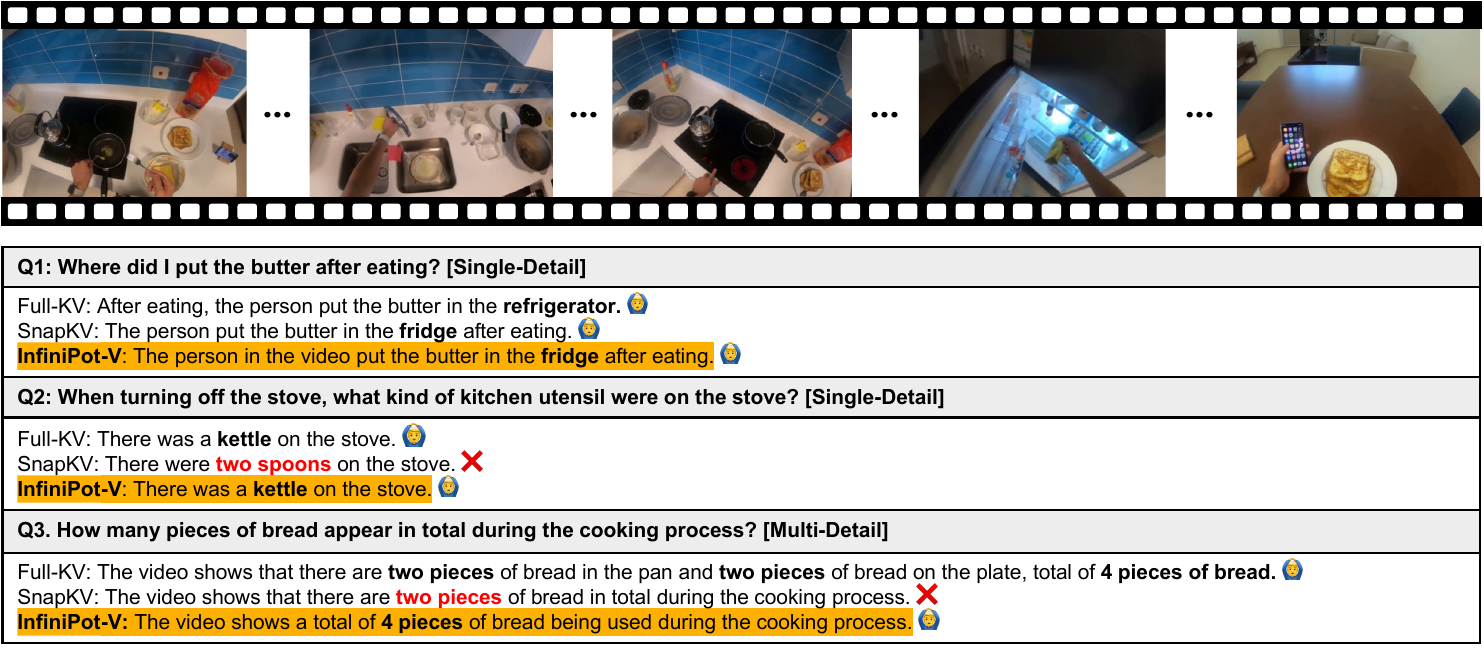}}
\caption{\textbf{Qualitative Results of Multi-Turn Conversation}: Full-KV uses 16K cache while InfiniPot-V and SnapKV employ 3K compressed KV cache. SnapKV performs query-guided cache compression based on Q1 before proceeding with multi-turn conversation. The video sample is from the MLVU ego reasoning subtask, using the \texttt{Qwen-2-VL-7B model}. 128 frame sampling is used.}\label{fig:multi_turn}
\vspace{-.2in}
\end{figure*}

\section{Multi-Turn Video Understanding Analysis}\label{appn:multi_turn}
Fig.~\ref{fig:multi_turn} presents a qualitative comparison between query-dependent (SnapKV)\cite{li2024snapkv} and query-agnostic (InfiniPot-V) KV cache compression approaches in multi-turn conversations with streaming video input. When SnapKV performs compression based on Q1, it generates answers almost identical to the Full KV cache for that specific query (Q1), answering that the butter was placed in the refrigerator. However, this query-guided compression strategy reveals significant limitations when handling different types of queries (Q2, Q3) about the same video content. Specifically, SnapKV makes critical errors in subsequent queries - misidentifying a kettle as "two spoons" in Q2 and incorrectly counting the total number of bread pieces in Q3.

In contrast, InfiniPot-V maintains accurate answers consistently across all three queries using the same 3K compressed KV cache. It correctly identifies that the butter was placed in the fridge (Q1), recognizes the kettle on the stove (Q2), and counts all 4 pieces of bread throughout the cooking process (Q3), demonstrating the effectiveness of query-agnostic compression for multi-turn streaming video scenarios.

\section{Why Query-Agnostic KV Cache Compression Matters for SVU?}\label{appn:query_agnostic}

In this section, we provide a detailed analysis of why query-agnostic compression is essential for Streaming Video Understanding (SVU), building upon the requirements discussed in Sec.~\ref{sec:background}. To demonstrate how these SVU-specific constraints impact existing KV cache compression methods, we present a case study across three representative scenarios.


\subsection{Preliminary: Attention-based KV Cache Compression}\label{appn:prelim}

Eviction-based KV cache compression reduces cache size by removing tokens with the lowest importance scores.
Employing attention scores for computing token importance scores is the predominant approach in previous methods~\cite{li2024snapkv,chen2024image,fu2024headkv,hooper2024squeezed}.

In methods such as SnapKV~\cite{li2024snapkv}, the importance scores $u_t$ of a token $x_t$ are computed by aggregating attention scores from the last $w$ tokens (i.e., observation window) which contain the user instruction:
\vspace{-.1in}
\begin{equation}
u_t = \sum_{i=N-w}^{N} \text{Attn}(x_{i} \rightarrow x_{t}),
\label{eq:importance-future}
\end{equation}
where $N$ is the current sequence length.
Using these scores, the KV cache is compressed by retaining the top-$M$ tokens with the highest aggregated attention scores.
Here, $M$ defines the memory budget: $\mathcal{I} = \text{TopK}(u, M)$ and $u = [u_1, \cdots, u_N]$ indicates the importance scores of all tokens.
The compressed Key and Value caches are then formed by extracting tokens at indices $\mathcal{I}$:
\begin{equation}
\tilde{K} = K[:, \mathcal{I}, :], \quad \tilde{V} = V[:, \mathcal{I}, :]
\label{eq:cache-compression}
\end{equation}
where $K, V \in \mathbb{R}^{H \times N \times D}$ are the uncompressed Key and Value caches with $H$ heads, $N$ tokens, and per-head dimension $D$. 
This approach has two characteristics: (1) it requires computing the full KV cache for all tokens before compression, and (2) it requires the user query to be present at the end of the context. We refer to these approaches as \textit{query-guided} or \textit{attention-based} cache compression methods.\footnote{Throughout this paper, "query" refers to the user's instruction or question related to the given video.}

\begin{figure}
    \centering
    \includegraphics[width=1\linewidth]{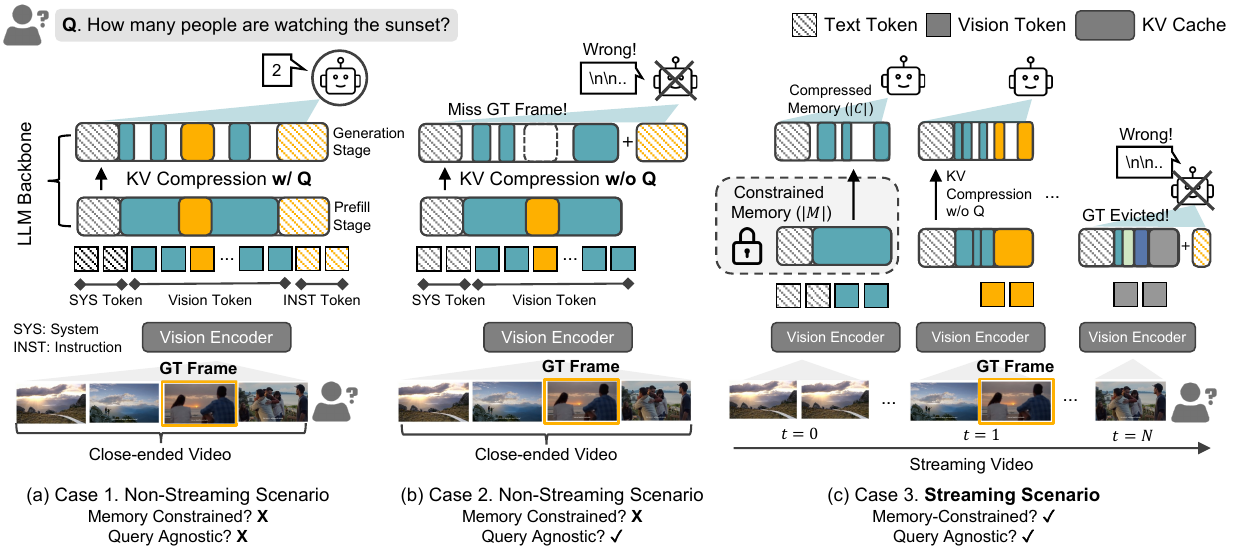}
\caption{\textbf{KV Cache Compression Case Study with SVU}: Illustration of cache control strategies under three conditions, differing in the presence of two core requirements for Streaming Video Understanding (SVU): memory constrained (MC) and query agnostic (QA). (a) Case 1: Query-guided compression retains relevant (GT) frames for accurate responses. (b) Case 2: Without query guidance, compression fails to preserve critical frames, resulting in inaccurate responses. (c) \textbf{Case 3} \textbf{(Streaming scenario)}: In streaming video processing, where frames arrive continuously, continual KV cache compression (CKV) is necessary, but queries are unavailable during compression.}
\label{fig:case_study}
\vspace{-2mm}
\end{figure}

\subsection{Case Study: Towards Streaming Video Understanding with CKV}\label{appn:case_study}

To investigate the applicability of attention-based KV cache compression methods to streaming video understanding, we examine three cache control strategies (Fig.~\ref{fig:case_study}).

\begin{wraptable}{r}{0.5\columnwidth}
\scriptsize
\begin{tabular}{c|c|c|cc@{}}
\toprule
Case & \makecell{$x_t$ Attention \\ Scoring} & \makecell{Prefill \\ $|M|$} & \makecell{Gen. \\ $|M|$ = 3K} & \makecell{Gen. \\ $|M|$ = 6K} \\ 
\midrule
Full KV & n/a & 25K & \multicolumn{2}{c}{68.75 (↑)} \\ \midrule
Case 1 & $\text{Attn}(q \rightarrow x_{t})$ & 25K & 68.01 & 68.40 \\ \midrule
\multirow{3}{*}{Case 2} & $\text{Attn}(q' \rightarrow x_{t})$ & \multirow{3}{*}{25K} & 60.35 & 63.42 \\ 
                        & $\text{Attn}(q'' \rightarrow x_{t})$ & & 60.60 & 63.50 \\ 
                        & $\text{Attn}(q_v \rightarrow x_{t})$ & & 60.32 & 62.28 \\ \midrule
\textbf{Case 3} & $\text{Attn}(q_v \rightarrow x_{t})$ & 3K/6K & 57.55 & 59.98 \\ 
\bottomrule
\end{tabular}
\caption{\textbf{Case study of Attention Scoring}: conducted on MLVU benchmark with \texttt{LLaVA-Next-Video-7B}. Note that memory-constrained setting (Case 3) shares the same budget during prefill and generation stages.}\label{tab:case_study}
\vspace{-.15in}

\end{wraptable}

\paragraph{Case 1.} Recent KV cache compression methods~\cite{li2024snapkv,fu2024headkv} assume full access to context and queries at compression time, as shown in Fig.~\ref{fig:case_study}(a). In this memory-unconstrained setting, the model observes the full input before compression. Previous works~\cite{li2024snapkv} have demonstrated that attention scores effectively identify query-relevant tokens KV cache (orange box corresponding to GT Frame in Fig.~\ref{fig:case_study}(a)), enabling compression that retains critical information while discarding less important tokens. As shown in Tab.~\ref{tab:case_study}, this approach maintains performance comparable to the uncompressed cache setup (68.01 vs 68.75) at the cost of large memory usage at compression, detailed in Fig.~\ref{fig:background}.

\paragraph{Case 2.} Fig.~\ref{fig:case_study}(b) illustrates how attention-based cache compression fails when user queries are unavailable during compression.
Under this scenario, although the memory budget is assumed unconstrained, the KV cache is compressed without consideration of (future) queries, causing important visual tokens (orange tokens cache corresponding to the GT Frame) lost during compression.
To quantify this degradation and explore alternatives, we test compression with generic queries ($q'$: "What is happening in this video?", $q'':$"What are the key events in this video?") and the last vision tokens ($q_v$) for importance scoring:

\begin{equation}
u^{\text{alt}}_t = \text{Attn}(q_{\text{alt}} \rightarrow x_t), \quad 
q_{\text{alt}} \in \{q', q'', q_v\}
\label{eq:alternative-importance}
\end{equation}

Tab.~\ref{tab:case_study} and Fig.~\ref{fig:case_study}(a) show that these alternatives significantly degrade performance (60.32 vs 68.75), even with unconstrained memory.

\paragraph{Case 3: Streaming Scenario.} Beyond the query-agnostic challenge in Case 2, deploying streaming video understanding on resource-constrained devices requires fixed memory usage for the KV cache.
For input video streams, these constraints necessitate continual compression when new frames arrive and memory capacity is reached, as shown in Fig.~\ref{fig:case_study}(c).
To evaluate this scenario, we use the query-agnostic approach from Case 2 with vision tokens ($q_v$) for importance scoring, while compressing the KV cache whenever memory limits are reached.
As shown in Tab.~\ref{tab:case_study}, this combined constraint further degrades performance (57.55 vs 60.32), highlighting the challenge of preserving key information under both query-agnostic and memory-constrained settings.

This case study reveals two key challenges for KV cache compression in streaming video: 1) the need for \textit{query-agnostic compression} due to continuous incoming video, and 2) the requirement to maintain \textit{fixed memory constraints}. These challenges cause significant performance drops in previous methods~\cite{li2024snapkv,kim-etal-2024-infinipot,chen2024image}, motivating Continual KV cache compression (CKV) specifically designed for memory-constrained streaming video.

\begin{table}[t]
  \centering
  \begin{minipage}[t]{0.5\textwidth}
    \vspace{0pt}                 
    \includegraphics[width=0.95\textwidth]{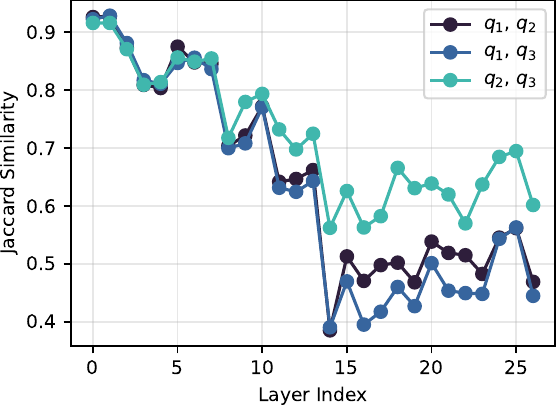}
    \vspace{0.5mm}
    \captionof{figure}{\textbf{Jaccard Similarity between KV Caches}: Compare KV cache sets selected by different queries ($q_1,q_2,q_3$) across layers.}
    \label{fig:jac_sim}
  \end{minipage}%
  \hfill
  \begin{minipage}[t]{0.45\textwidth}
    \vspace{2pt}
    \resizebox{\textwidth}{!}{%
      \begin{tabular}{c|cc|cc}
        \toprule
        Context & \multicolumn{2}{c|}{Case 1, 2} & \multicolumn{2}{c}{Case 3 (Ours)} \\
        Length & Mem (GB) & TTFT (s) & Mem (GB) & TTFT (s) \\ 
        \midrule
        5K    & 21.29 & 0.98 & 20.93 & 1.08 \\
        25K   & 33.76 & 1.21 & 21.60 & 1.12 \\
        50K   & 58.55 & 2.12 & 22.16 & 1.17 \\
        100K  & 79.38 & 3.27 & 22.85 & 1.20 \\
        \bottomrule
      \end{tabular}%
    }
    \vspace{2mm}
    \captionof{table}{\textbf{Peak GPU Memory and TTFT}: Comparison of peak memory usage and Time-To-First-Token(TTFT) across different context lengths for memory-unconstrained (Case 1, 2) and memory-constrained (Case 3) approaches.}
    \label{tab:memory_ttft}
  \end{minipage}
  \vspace{-.2in}
\end{table}

\paragraph{Attention Scoring Analysis} We further analyze the query-dependent nature of attention-based KV cache compression using the VideoMME benchmark. To investigate why performance varies with different queries, we compute the Jaccard similarity between token sets selected for different queries across layers using attention scores at each layer. For this analysis, $q_1, q_2, q_3$ represent three distinct questions associated with the same video sample in the VideoMME benchmark. As shown in Fig.~\ref{fig:jac_sim}, the similarity between token sets decreases significantly in the middle-to-late layers, dropping to around 0.4. This indicates that each query selects a different set of tokens, particularly in deeper layers. This analysis highlights that attention-based scoring methods inherently select query-specific tokens, explaining the performance degradation when query information is unavailable or changes during streaming video scenarios.

\section{Memory and Latency Measurement Results}\label{appn:memory_latency}

Table~\ref{tab:memory_ttft} presents measurements of peak memory consumption and Time-To-First-Token (TTFT) during the prefill stage, conducted on a single NVIDIA A100-80GB GPU using PyTorch. The experiments averaged over five runs with three warmup iterations, compare the performance of memory-unconstrained (Case 1, 2) and memory-constrained (Case 3) approaches across various context lengths. For memory-unconstrained methods, we observe a linear growth in memory requirements, escalating from 21.29 GB at 5K tokens to 79.38 GB at 100K tokens, accompanied by a proportional increase in TTFT from 0.98 to 3.27 seconds.

Our memory-constrained continual KV cache compression (Case 3) exhibits remarkably different behavior. Despite the increasing context length, the peak memory usage shows only minimal growth, rising modestly from 20.93 GB at 5K tokens to 22.85 GB at 100K tokens. Similarly, the TTFT remains relatively stable, increasing from 1.08 to 1.20 seconds across the same range. These detailed measurements demonstrate that our approach effectively maintains near-constant resource utilization while processing extended video frames.

\section{Related Work}\label{sec:related}

\subsection{MLLMs for Long Video Understanding}
Recent advances in long-context MLLMs have attracted significant attention. Notable examples include Gemini-2.0~\cite{google2024cgemini-2.0}, supporting streaming video; LongVILA~\cite{Chen2025longvila}, capable of handling up to 6,000 video frames; LLaVA-Next-Video~\cite{zhang2024llavanext-video}, which leverages high-quality synthetic instruction data; and Qwen-2-VL~\cite{Qwen2VL}, enabling hour-long video analysis via multimodal RoPE.

\subsection{Input-Vision Compression (IVC)}
To address the computational demands of long-form video processing, several approaches have been proposed to compress redundant visual information before it enters the backbone LLM. 

LongVU~\cite{shen2024longvu} adopts query-dependent input frame sampling and redundant pixel removal for fine-grained video understanding, but the two-tower vision encoding results in high latency during input sampling, making it impractical for streaming scenarios. Additionally, this approach requires training specialized models to operate in the proposed manner, limiting its applicability to existing pre-trained models. 

DyCoke~\cite{tao2024dycokedynamiccompressiontokens} reduce redundancies between adjacent frames at the input video level and dynamically updates query-related tokens in the KV cache from external storage. Slow-Fast-LLaVA-1.5~\cite{xu2025slowfastllava15familytokenefficientvideo} proposes dividing input video processing into separate slow and fast pathways, using different projection methods to reduce input vision tokens. However, this approach still suffers from the limitation of requiring all input vision tokens to be processed simultaneously and necessitates additional model training.

\subsection{KV Cache Compression (KVC)}
Understanding the long context in MLLMs demands efficient KV cache control to manage memory growth and latency overhead. KV cache compression methods can be broadly categorized into query-dependent and query-agnostic approaches.

\paragraph{Query-Dependent KV cache Compression.}
Methods like SnapKV~\cite{li2024snapkv}, H2O~\cite{zhang2023ho}, HeadKV~\cite{fu2024headkv} and ThinK~\cite{xu2025think} leverage query-to-context attention scores to identify crucial KV entries but require the full context to be prefilled before compression, making them impractical under memory constraints. In the multimodal domain, FastV~\cite{chen2024image} accelerates prefill by pruning vision tokens at certain layers based on their attention scores from the final query token.
SparseVLM~\cite{zhang2024sparsevlm} selects visual tokens relevant to user queries via cross-attention. Overall, query-dependent methods effectively compress context but struggle to handle diverse queries for the given context after compression~\cite{tang2025razorattention}. ReKV~\cite{Shangzhe2025streaming} addresses streaming video scenarios by offloading video-related KV cache to CPU memory and retrieving query-dependent cache entries on demand. This approach relies on external storage and suffers from data transfer overhead, making it unsuitable for memory-constrained streaming video understanding.

\paragraph{Query-Agnostic KV cache Compression.}
Recent works pursue query-agnostic KV cache compression to eliminate reliance on future queries~\cite{ge2024model,devoto2024simple,hooper2024squeezed, kim2025kvzipqueryagnostickvcache, corallo2025ragtaskawarekvcache,park2025keydiffkeysimilaritybasedkv}.
In particular, SqueezedAttention~\cite{hooper2024squeezed} uses key-based clustering but requires full-context encoding, limiting its applicability to memory-constrained settings.
InfiniPot~\cite{kim-etal-2024-infinipot} compresses context by approximating potential user queries through a task-specific proxy prompt, but it's fixed prompt restricts flexibility. In the vision domain, HiRED~\cite{arif2024hired} and FasterVLM~\cite{zhang2024fastervlm} utilize [CLS] token attention scores for compression decisions. However, their reliance on special tokens restricts their application to recent MLLMs that lack such tokens~\cite{qwen2.5,zhang2024llavanext-video}, limiting their broader applicability.

\section{Experimental Results Data}

\begin{table*}[t!]
    \centering
    \resizebox{\textwidth}{!}
    {
    \begin{tabular}{c|cc|c|c|c|cccc|cccc|c|c}
    \toprule
    \multirow{2}{*}{\rotatebox[origin=c]{90}{Case}}& \multicolumn{2}{c|}{Streaming} & Compression & Prefill & Decoding & \multicolumn{4}{c|}{VideoMME} & \multicolumn{4}{c|}{MLVU} & \multirow{2}{*}{LVB} & \multirow{2}{*}{Avg.} \\
    & MC & QA & Method & Budget & Budget & Short & Medium & Long & Avg. & Holistic & Single & Multi & Avg. & & \\
    \midrule
    \multicolumn{16}{c}{\textbf{Qwen-2-VL-7B}} \\
    \midrule
    & - & - & Full KV & 50K & 50K & 74.68 & 62.11 & 55.00 & 63.93 & 76.34 & 73.91 & 43.29 & 65.85 & 58.77 & 62.85 \\
    \cmidrule{1-16}
    \multirow{4}{*}{\rotatebox[origin=c]{90}{Case 1}} & \multirow{4}{*}{\ding{55}} & \multirow{4}{*}{\ding{55}} & FastV~\cite{chen2024image} & \multicolumn{2}{c|}{48/3K~($R=2.8$)} & 54.11 & 50.11 & 48.67 & 50.96 & 69.59 & 59.40 & 33.84 & 55.01 & 47.94 & 51.30 \\
    & & & ($L=2$) & \multicolumn{2}{c|}{48/6K~($R=5.8$)} & 59.67 & 54.55 & 50.78 & 55.00 & 72.00 & 64.08 & 33.47 & 57.60 & 50.53 & 54.38 \\
    \cmidrule{4-16}
    & & & \multirow{2}{*}{SnapKV~\cite{li2024snapkv}} & 50K & 3K & 74.00 & 61.00 & 54.22 & 63.07 & 77.08 & 67.49 & 39.07 & 62.11 & 59.06 & 61.42 \\
    & & & & 50K & 6K & 74.22 & 60.55 & 54.33 & 63.03 & 77.59 & 73.91 & 42.90 & 66.10 & 58.80 & 62.64 \\
    \midrule
    \multirow{4}{*}{\rotatebox[origin=c]{90}{Case 2}} & \multirow{4}{*}{\ding{55}} & \multirow{4}{*}{\ding{51}} & \multirow{2}{*}{Uniform} & 50K & 3K & 70.33 & 54.67 & 49.55 & 58.18 & 72.29 & 59.06 & 33.51 & 55.54 & 59.80 & 57.84 \\
    & & & & 50K & 6K & 72.00 & 58.78 & 52.11 & 60.96 & 77.08 & 67.49 & 39.07 & 62.11 & 59.11 & 60.73 \\
    \cmidrule{4-16}
    & & & \multirow{2}{*}{SnapKV$^\dag$} & 50K & 3K & 69.00 & 54.00 & 50.67 & 57.89 & 75.88 & 63.48 & 35.35 & 58.99 & 56.70 & 57.86 \\
    & & & & 50K & 6K & 72.11 & 57.56 & 52.22 & 60.63 & 76.46 & 66.43 & 36.22 & 60.66 & 56.72 & 59.34 \\
    \midrule
    \multirow{16}{*}{\rotatebox[origin=c]{90}{Case 3 (CKV)}} & \multirow{16}{*}{\ding{51}} & \multirow{16}{*}{\ding{51}} & \multirow{4}{*}{Uniform} & 3K & 3K & 66.00 & 52.44 & 48.00 & 55.48 & 72.54 & 59.00 & 33.51 & 55.59 & 55.21 & 55.43 \\
    & & & & 6K & 6K & 72.33 & 53.33 & 48.67 & 58.11 & 72.55 & 62.19 & 33.67 & 57.00 & 55.82 & 56.98 \\
    & & & & 12K & 12K & 74.00 & 55.33 & 51.44 & 60.26 & 75.94 & 65.53 & 37.01 & 60.36 & 57.91 & 59.51 \\
    & & & & 24K & 24K & 74.22 & 59.22 & 53.22 & 62.22 & 77.22 & 71.10 & 40.78 & 64.18 & 58.60 & 61.67 \\
    \cmidrule{4-16}
    
    & & & \multirow{4}{*}{SnapKV$^\ddag$} & 3K & 3K & 66.67 & 52.22 & 49.89 & 56.26 & 75.88 & 63.48 & 35.35 & 58.99 & 54.91 & 56.72 \\
    & & & & 6K & 6K & 72.00 & 55.33 & 51.33 & 59.55 & 76.46 & 66.43 & 36.22 & 60.66 & 55.15 & 58.45 \\
    & & & & 12K & 12K & 74.44 & 58.89 & 52.89 & 62.07 & 75.71 & 68.61 & 35.98 & 61.31 & 56.89 & 60.09 \\
    & & & & 24K & 24K & 74.22 & 61.00 & 53.78 & 63.00 & 77.66 & 71.82 & 39.90 & 64.37 & 59.09 & 62.15 \\
    \cmidrule{4-16}
    
    & & & \multirow{4}{*}{InfiniPot~\cite{kim-etal-2024-infinipot}}& 3K & 3K & 67.11 & 54.55 & 51.00 & 57.55 & 74.94 & 61.80 & 36.60 & 58.36 & 54.00 & 56.64 \\
    & & & & 6K & 6K & 72.89 & 57.33 & 51.33 & 60.52 & 75.02 & 63.18 & 37.09 & 59.11 & 54.64 & 58.09 \\
    & & & & 12K & 12K & 74.00 & 57.78 & 53.22 & 61.67 & 74.46 & 66.46 & 38.30 & 60.70 & 56.94 & 59.77 \\
    & & & & 24K & 24K & 74.22 & 60.55 & 53.56 & 62.78 & 76.03 & 71.11 & 40.29 & 63.71 & 57.85 & 61.44 \\
    
    \cmidrule{4-16}
    \mscellf & & & & 3K & 3K & 73.89 & 57.78 & 51.78 & 61.11 & 77.73 & 70.38 & 43.15 & 64.70 & 57.64 & 61.15 \\
    \mscells & & & & 6K & 6K & 74.11 & 60.78 & 53.44 & 62.78 & 77.16 & 72.31 & 44.75 & 65.82 & 58.40 & 62.33 \\
    \mscellt & & & & 12K & 12K & 74.22 & 62.68 & 53.89 & 63.59 & 76.90 & 73.41 & 43.97 & 65.99 & 59.18 & 62.92 \\
    \mscellfo & & & \multirow{-4}{*}{\textbf{InfiniPot-V}} & 24K & 24K & 74.22 & 63.22 & 53.11 & 63.52 & 76.91 & 73.97 & 42.18 & 65.73 & 58.94 & 62.73 \\
    
    \midrule
    \multicolumn{16}{c}{\textbf{LLaVA-Next-Video-7B}} \\
    \midrule
    & - & - & Full KV & 25K & 25K & 74.33 & 60.11 & 54.11 & 62.85 & 80.60 & 73.73 & 49.43 & 68.75 & 63.55 & 65.05 \\
    \midrule
    \multirow{4}{*}{\rotatebox[origin=c]{90}{Case 1}} & \multirow{4}{*}{\ding{55}} & \multirow{4}{*}{\ding{55}} & \multirow{2}{*}{Uniform} & 25K & 3K & 74.33 & 62.33 & 55.00 & 63.89 & 80.29 & 72.38 & 49.19 & 68.01 & 62.35 & 64.75 \\
    & & & & 25K & 6K & 73.89 & 62.00 & 54.78 & 63.56 & 80.66 & 72.25 & 49.62 & 68.19 & 62.55 & 64.76 \\
    \cmidrule{4-16}
    & & & \multirow{2}{*}{SnapKV~\cite{li2024snapkv}} & 25K & 3K & 74.44 & 59.89 & 53.78 & 62.70 & 80.41 & 73.01 & 49.67 & 68.46 & 62.34 & 64.50 \\
    & & & & 25K & 6K & 74.44 & 60.11 & 53.78 & 62.78 & 80.60 & 73.45 & 49.48 & 68.64 & 62.34 & 64.59 \\
    \midrule
    \multirow{4}{*}{\rotatebox[origin=c]{90}{Case 2}} & \multirow{4}{*}{\ding{55}} & \multirow{4}{*}{\ding{51}} & \multirow{2}{*}{Uniform} & 25K & 3K & 66.33 & 54.00 & 49.67 & 56.67 & 75.12 & 59.65 & 38.55 & 58.04 & 59.14 & 57.95 \\
    & & & & 25K & 6K & 71.00 & 56.33 & 51.55 & 59.63 & 77.84 & 65.60 & 43.92 & 62.90 & 61.69 & 61.41 \\
    \cmidrule{4-16}
    & & & \multirow{2}{*}{SnapKV$^\dag$} & 25K & 3K & 64.00 & 54.55 & 51.11 & 56.55 & 78.53 & 59.73 & 41.69 & 59.94 & 56.19 & 57.56 \\
    & & & & 25K & 6K & 69.55 & 58.44 & 52.78 & 60.26 & 80.86 & 63.65 & 45.07 & 63.26 & 59.90 & 61.14 \\
    \midrule
    

    \multirow{16}{*}{\rotatebox[origin=c]{90}{Case 3 (CKV)}} & \multirow{16}{*}{\ding{51}} & \multirow{16}{*}{\ding{51}} & \multirow{4}{*}{Uniform} & 1.5K & 1.5K & 56.22 & 46.89 & 44.00 & 49.04 & 69.72 & 52.53 & 36.53 & 52.87 & 54.92 & 52.28 \\
    & & & & 3K & 3K & 59.22 & 51.55 & 47.44 & 52.74 & 74.30 & 57.25 & 36.48 & 56.19 & 54.40 & 54.44 \\
    & & & & 6K & 6K & 64.89 & 55.67 & 49.78 & 56.78 & 76.71 & 61.14 & 34.55 & 57.99 & 57.72 & 57.50 \\
    & & & & 12K & 12K & 72.67 & 59.89 & 53.00 & 61.85 & 80.03 & 67.33 & 44.31 & 64.38 & 61.04 & 62.42 \\
    \cmidrule{4-16}

    & & & \multirow{4}{*}{SnapKV$^\ddag$} & 1.5K & 1.5K & 52.40 & 58.00 & 51.33 & 47.89 & 74.92 & 56.89 & 32.62 & 55.11 & 53.65 & 52.22 \\
    & & & & 3K & 3K & 62.11 & 54.55 & 48.55 & 55.07 & 76.94 & 59.18 & 35.71 & 57.55 & 54.71 & 55.78 \\
    & & & & 6K & 6K & 66.33 & 56.11 & 51.11 & 57.85 & 79.60 & 62.15 & 37.12 & 59.98 & 57.81 & 58.55 \\
    & & & & 12K & 12K & 72.11 & 58.00 & 53.11 & 61.07 & 79.71 & 67.99 & 44.89 & 64.74 & 58.83 & 61.55 \\
    \cmidrule{4-16}
    
    & & & \multirow{4}{*}{InfiniPot~\cite{kim-etal-2024-infinipot}} & 1.5K & 1.5K & 53.22 & 51.11 & 47.55 & 53.11 & 69.89 & 56.44 & 30.54 & 52.88 & 52.14 & 52.71 \\
    & & & & 3K & 3K & 58.22 & 51.78 & 49.33 & 54.22 & 72.42 & 55.88 & 34.45 & 54.48 & 52.43 & 53.71 \\
    & & & & 6K & 6K & 62.44 & 53.89 & 51.11 & 55.81 & 76.46 & 57.97 & 37.07 & 57.28 & 55.58 & 56.22 \\
    & & & & 12K & 12K & 70.55 & 59.22 & 52.55 & 60.77 & 79.84 & 67.81 & 45.57 & 64.89 & 59.23 & 61.63 \\
    \cmidrule{4-16}
    
     \mscellf & & & & 1.5K & 1.5K & 63.89 & 52.55 & 47.11 & 54.52 & 77.08 & 57.32 & 34.64 & 56.49 & 56.48 & 55.83 \\
     \mscells & & & & 3K & 3K & 67.78 & 56.22 & 50.33 & 58.11 & 77.88 & 65.74 & 40.31 & 61.94 & 58.37 & 59.47 \\
     \mscellt & & & & 6K & 6K & 72.44 & 59.55 & 51.33 & 61.11 & 80.03 & 69.41 & 43.93 & 65.16 & 60.86 & 62.38 \\
     \mscellfo & & & \multirow{-4}{*}{\textbf{InfiniPot-V}}  & 12K & 12K & 73.89 & 58.67 & 52.11 & 61.55 & 80.91 & 71.16 & 51.57 & 68.35 & 61.84 & 63.91 \\ 
    \bottomrule
    \end{tabular}
    }
    \caption{\textbf{InfiniPot-V vs KVC} Offline long video understanding evaluation results under memory-constrained scenario (case 3), with MC (Memory-Constrained) and QA (Query-Agnostic) conditions marked. Results are reported on (1) Video-MME - Short: -3min, Medium: 3-30min, Long: 30min-2h, (2) MLVU - Holistic, Single-Detail, Multi-Detail LVU, and (3) LVB (LongVideoBenchmark).}\label{tab:full_table}
    \vspace{-0.2in}
\end{table*}

\begin{table*}[t]
\centering
\resizebox{\textwidth}{!}{
\begin{tabular}{lccccccccccc}
\toprule
Qwen-2-VL & Vision & Decoding & \multicolumn{4}{c}{MLVU} & \multicolumn{4}{c}{VideoMME} & \\
\cmidrule(lr){4-7} \cmidrule(lr){8-11}
IVC Methods & Budget & Budget & Holistic & Single & Multi & Avg. & Short & Med & Long & Avg. & Avg. \\
\midrule
Full KV        & 50K & 50K & 76.3 & 73.9 & 43.3 & 65.9 & 74.7 & 62.1 & 55.0 & 63.9 & 64.2 \\ \midrule
Uniform     & 50K & 6K  & 77.7 & 69.8 & 41.6 & 64.0 & 74.9 & 58.0 & 52.8 & 61.9 & 62.5 \\
TTM~\cite{tao2024dycokedynamiccompressiontokens} & 50K & 6K  & 78.2 & 70.0 & 42.7 & 64.5 & 74.9 & 59.2 & 52.7 & 62.3 & 62.9 \\
STC~\cite{shen2024longvu} & 50K & 6K  & 77.9 & 71.5 & 44.7 & 65.7 & 74.3 & 59.6 & 54.6 & \textbf{62.8} & \textbf{63.8} \\
\cmidrule(lr){1-12}
\mscellt \textbf{InfiniPot-V}        & 6K  & 6K  & 77.2 & 72.3 & 44.7 & \textbf{65.8} & 74.1 & 60.8 & 53.4 & \textbf{62.8} & \textbf{63.8} \\ \midrule
Uniform     & 50K & 3K  & 75.7 & 66.5 & 38.6 & 61.1 & 72.2 & 53.4 & 50.0 & 58.6 & 59.4 \\
TTM~\cite{tao2024dycokedynamiccompressiontokens}         & 50K & 3K  & 77.3 & 67.8 & 39.5 & 62.4 & 72.7 & 56.2 & 52.2 & 60.4 & 61.0 \\
STC~\cite{shen2024longvu} & 50K & 3K  & 76.9 & 68.2 & 41.7 & 63.1 & 71.2 & 55.9 & 53.7 & 60.3 & 61.3 \\
\cmidrule(lr){1-12}
\mscellt \textbf{InfiniPot-V}    & 3K  & 3K  & 77.7 & 70.4 & 43.2 & \textbf{64.7} & 73.9 & 57.8 & 51.8 & \textbf{61.1} & \textbf{62.5} \\
\bottomrule
\end{tabular}
}
\vspace{1mm}
\caption{\textbf{InfiniPot-V vs IVC}: Performance comparison between Input-Vision Compression (IVC) methodology and InfiniPot-V. Vision budget denotes the vision token length before IVC, while decoding budget refers to the input token length used during decoding. Evaluated using Qwen-2-VL with MLVU and VideoMME datasets.}
\label{tab:full_ivc}
\end{table*}

\subsection{Comparison between InfiniPot-V and KVC}

Tab.~\ref{tab:full_table} provides a detailed performance comparison between KV cache compression (KVC) methods and InfiniPot-V across offline video understanding (OVU) benchmarks under various compression ratios for two models: Qwen-2-VL and LLaVA-Next.

In Case 1, where the full prefill is conducted and the final query is accessible at compression time, FastV demonstrates significantly inferior performance at similar compression ratios due to its aggressive token-pruning strategy. In contrast, SnapKV shows robust performance at high compression ratios across both models by utilizing the full context KV cache and retaining vision tokens that are highly correlated with the given query.

Case 2 examines the query-agnostic setting, where, as explored in our earlier case study in Appendix.~\ref{appn:case_study}, SnapKV exhibits notable performance degradation across both models when applied in a query-agnostic manner, showing performance comparable to uniform selection baseline.

In Case 3, which represents the CKV framework scenario where the constrained memory budget is used for both prefill and decoding stages, InfiniPot-V significantly outperforms all three baselines across various compression ratios on both models, as showcased in Fig.~\ref{fig:cache_compression}.

\subsection{Comparison between InfiniPot-V and IVC}
Table~\ref{tab:full_ivc} presents a performance comparison between Input-Vision Compression (IVC) methods and InfiniPot-V on the MLVU and Video-MME benchmarks using the Qwen-2-VL model. Under a 6K decoding budget, the IVC methods demonstrate robust overall performance by utilizing the full vision encoding budget (50K tokens). InfiniPot-V achieves comparable or slightly superior performance to these methods while operating under constrained memory budgets for both vision encoding and decoding stages (6K tokens).

When the decoding budget is compressed to 3K tokens, the IVC methods exhibit performance degradation, with LongVU's STC methodology achieving the highest performance among the IVC approaches. Notably, InfiniPot-V demonstrates both efficiency and effectiveness by achieving higher accuracy than IVC methods that utilize the full vision encoding budget, while operating under constrained budgets (3K) for both vision encoding and decoding stages. 

\section{Limitation and Future Work}

InfiniPot-V introduces the first training-free, query-agnostic framework for memory-constrained streaming video understanding, enabling length-independent KV cache compression with minimal accuracy loss across long-form, real-time scenarios. However, several avenues exist for further advancement. Current approaches focus primarily on vision tokens, yet real-world streaming applications involve multiple modalities including speech, text, and video simultaneously. Future work could extend our framework to unified multimodal compression, enabling more realistic and comprehensive streaming understanding systems that efficiently manage diverse input types within fixed memory constraints.

Additionally, our current fixed budget allocation between TaR and VaN components could benefit from adaptive mechanisms that dynamically adjust compression ratios based on input characteristics—allocating more resources to temporal redundancy reduction for static scenes or prioritizing spatial importance for content-rich frames. Furthermore, while InfiniPot-V's training-free nature ensures broad applicability, end-to-end learning approaches could optimize models specifically for continual compression scenarios, potentially enabling more aggressive compression ratios through learned token importance estimation~\cite{huang2025locretenhancingevictionlongcontext} tailored to streaming video understanding tasks.


\end{document}